\documentclass{openjournal}

\usepackage{lipsum}

\usepackage{xcolor}
\usepackage{textgreek}
\usepackage[utf8]{inputenc}
\usepackage[english]{babel}

\usepackage{hyperref}
\hypersetup{
    unicode, 
    colorlinks=true,
    linkcolor=linkcolor,
    citecolor=linkcolor,
    filecolor=linkcolor,
    urlcolor=linkcolor,
}
\usepackage{color,colortbl}
\definecolor{linkcolor}{rgb}{0.0,0.3,0.5}
\usepackage{tensind}
\tensordelimiter{?}
\DeclareGraphicsExtensions{.bmp,.png,.jpg,.pdf}
\usepackage{verbatim}
\usepackage[normalem]{ulem}
\usepackage{orcidlink}
\usepackage{soul}

\graphicspath{ {./figs/} }

\begin{document}
\title{Exploring the MBTI Distribution Among Chinese Undergraduate Physics Students: The Influence of Family Income on Career Trajectories}

\author{Songyang Bai\orcidlink{0009-0003-4351-8483}}
\email{songyang.bai@mail.utoronto.ca}
\affiliation{Department of Physics, University of Toronto}

\author{Weitian Chen\orcidlink{0000-0001-9457-4694}}
\email{weitian.chen@mail.utoronto.ca}
\affiliation{Department of Physics, University of Toronto}

\author{Zihan Gao}
\email{zgao13@ualberta.ca}
\affiliation{Department of Statistics, University of Alberta}

\begin{abstract}
This study investigated the distribution of MBTI personality types among physics undergraduates at Zhejiang University and analyzed their career aspirations, family income and mental health status. A comprehensive survey of 68 questions, including 52 multiple-choice questions and 16 short-answer questions, assessed various personality traits and their correlation with academic intent and emotional state. The results showed that INTJ and INTP personality types showed the most significant tendency to pursue academic research. They usually come from middle and above-class backgrounds. Notably, these two personality types accounted for about 41\% of the total surveyed population, while NT types combined accounted for about 58\%. This indicates that although NT personality is more suitable for academic research, students with introverted tendencies are more suitable. Research has further shown that people with ISTJ personalities often exhibit unexpectedly strong academic interests. In addition, individuals who aspire to enter academia generally maintain a stable state of mind, as evidenced by their consistent work efficiency in the face of personal challenges. Interestingly, although some aspiring academics come from financially stable families, most have encountered financial constraints. These results contribute to a deeper understanding of how personality traits and socioeconomic factors influence the academic career paths of physics students.
\end{abstract}

\maketitle

\section{Introduction}
\label{sec:intro}
The relationship between personality traits, socioeconomic factors, and career aspirations has been the focus of research in education and psychology. The Myers-Briggs Type Indicator (MBTI) is widely used to assess individual differences in behavior and decision-making styles under different personality frameworks \citep{myers1998}. This study examines the distribution of MBTI types among physics undergraduates at Zhejiang University to explore the correlation between MBTI types and students' academic ambitions, family income, and mental health.

Recent research has emphasized the importance of personality traits in shaping career paths, particularly in demanding fields such as academia and scientific research \citep{buss2019evolutionary}. Studies indicate that individuals with specific MBTI types, such as INTJs and INTPs, are more inclined to pursue knowledge and truth. These individuals typically demonstrate stronger problem-solving skills, which are essential in academic settings \citep{mccrae2003}. Moreover, socioeconomic background plays a critical role in determining access to educational resources and influencing career trajectories \citep{sirin2005socioeconomic}. However, further exploration is required to understand the interaction between socioeconomic factors and personality types.

The current study aims to bridge this gap by investigating a diverse group of physics undergraduates. To investigate the significant influence of personality traits on students' willingness to engage in academic research and the influence of socioeconomic factors, especially family income, on these aspirations, to predict what kind of personality and family environment are more suitable for students to enter the academic field. A 68-question questionnaire assessed various aspects of the student's personality, career intentions and emotional health. The findings not only contribute to understanding the applicability of MBTI in educational contexts but also provide insight into the socioeconomic dynamics that influence the career choices of Chinese physics undergraduates. This study aims to provide a comprehensive analysis of how personality types and socioeconomic factors interact to shape the academic and career aspirations of future scientists and the potential implications for educators, policymakers, and academic institutions regarding the importance of supporting diverse student backgrounds in promoting a thriving academic community.

\section{Methods}
\subsection{participants}
Three hundred physics undergraduates from Zhejiang University were recruited to participate in the investigation. The Department of Physics at Zhejiang University is one of the top physics majors in China. The results of academic intentions and family income obtained through such institutions are more valuable for reference. About 25 percent of this cohort's participants were women, and about 75 percent were men. The age range of the participants was 17-23 years old, and the overall age distribution was relatively uniform, ensuring the homogeneity and representativeness of the samples, which was convenient for subsequent analysis. To enhance the broad applicability of the study, the selection of participants included students of different academic levels, of which 20\% were first-year students, 25\% were second-year students, 30\% were third-year students, and 25\% were fourth-year students. This grade distribution can fully reflect the differences in personality characteristics and economic background of students at different academic levels.

All participants participated in the study voluntarily and provided informed consent before filling out the questionnaire to ensure voluntary participation and confidentiality of the information. Emphasize the voluntary and anonymous nature of participation in the recruitment process to enhance trust and willingness among participants. In addition, we recruited undergraduate students who were studying or had completed more than one semester of physics courses to ensure that they had a certain level of knowledge and foundation in the discipline, thus making the survey results more valid and reliable. Participants from different social and economic backgrounds were recruited to improve the generality of the findings, and were designed to cover students from all family situations.
\subsection{procedure}
In this study, a questionnaire survey was used to collect data. The research team developed a questionnaire containing multiple-choice, grading questions, and simple essay questions to assess participants' MBTI type, family income, and academic achievement. The questionnaire underwent three rounds of testing and revisions to ensure the clarity and relevance of the questions. The final version consists of three main sections: basic information on the participants (gender, age, family income), MBTI personality type assessment, academic achievement, mental health status, and self-reports of related learning attitudes. The questionnaire was conducted online, and participants estimated the completion time to be 10-15 minutes.During the data collection process, the research team distributed questionnaire links to participants via email and social media, and prizes were distributed to participants. Posters are also displayed in the public areas of the Physics Department to increase participation. Throughout the survey in the spring semester of 2024, the research team provided technical support to address any questions participants encountered while completing the questionnaire. In addition, to ensure the smooth conduct of the survey, we also actively communicate with participants to answer their questions and enhance their trust in the research.
\subsection{Measurement tools}
In this study, the MBTI scale based on Jungian personality theory was used to evaluate individual preferences from four dimensions: extroversion-introversion (E-I), sense-intuition (S-N), thought-feeling (T-F) and judge-perception (J-P). The MBTI type is determined by a combination of four binary dimensions, reflecting an individual's tendencies in information processing, decision-making, and lifestyle. Specifically, the characteristics of each type can be further illuminated by eight cognitive functions, including extraverted intuition (Ne), introverted intuition (Ni), extraverted thinking (Te), introverted thinking (Ti), extraverted sense (Se), introverted sense (Si), extraverted sense (Fe), and introverted sense (Fi). For example, the INTJ personality is mainly composed of introverted intuition (Ni), extroverted thinking (Te), introverted feeling (Fi), and extroverted feeling (Se). This unique combination gives them a unique way of thinking, making decisions and solving problems \citep{mccormack1990mmpi}.

By selecting the corresponding description option, participants were assigned a four-letter type code that represented their personality traits. Previous studies have shown that MBTI has high internal consistency confidence in large samples, exceeding 0.90 \citep{jung1971personality}, indicating high reliability and validity in assessing individual personality types. To measure household income levels, we used multiple-choice questions that asked participants to choose the option that best suited their financial situation accurately. Specifically, income levels are divided into five groups: above ¥20,000, ¥10,000-20,000, ¥6,000-10,000, ¥3,000-6,000, and below ¥3,000. This systematic classification is designed to fully reflect the economic background of the participants and provide detailed income information for subsequent analysis.

In terms of academic achievement, we obtained the academic grades of the participants through self-reporting and selected the options as A (outstanding performance), B (good performance), C(adequate performance), D (marginal performance), and F (Inadequate performance) to reflect their performance in the learning process. In addition, participants were asked to rate a series of statements based on how they felt on a scale of -10 to 10 (-10 being "strongly disagree," 0 being "neutral," and 10 being "strongly agree"). These statements relate to various external and internal factors that affect their academic performance, such as "\textit{I believe that financial inadequacies have significantly impacted my motivation to study}. Using this scoring method, the team could quantitatively analyze participants' attitudes and feelings. In order to ensure the clarity and validity of the questionnaire, the questionnaire was designed through a preliminary small-scale pilot study, and the questionnaire was adjusted according to the participants' feedback. The final version of the questionnaire includes personality traits, family income, academic performance and other factors that may affect academic performance, such as study habits, mental state, and family support (the full questionnaire is in the appendix).

\subsection{data analysis tools}
This study adopts a quantitative research approach and is structured as a cross-sectional investigation.  The selection of samples relies on convenience sampling, and the sample size comprises 300 students from the Department of Physics of Zhejiang University.  By analyzing this group, we aim to explore the interrelationship among MBTI type, family income, and academic achievement.  During data analysis, descriptive statistical methods, including the mean value, standard deviation, frequency, and percentage, were initially utilized to comprehensively depict the essential characteristics of the participants comprehensively.  This aspect provides fundamental data support for subsequent analyses and facilitates understanding of the overall composition of the sample.  To explore the relationship between MBTI type, family income, and academic achievement, we employed the Chi-square Test to evaluate the independence between MBTI type and family income level. We utilized one-way analysis of variance (ANOVA) to examine the differences in academic achievement among different MBTI types.  Specifically, we conducted multivariate ANOVA with household income level as the independent variable, academic achievement as the dependent variable, and MBTI type as the grouping variable.  This methodology can effectively identify correlations between variables and furnish statistically significant results.

Data analysis was conducted using SPSS statistical software, and the significance level of all statistical tests was set at 0.05.    A pie chart was employed to visualize the distribution of personality types, illustrating the proportions of various MBTI types among the participants. Additionally, pie charts were utilized to present the academic performance of different personality types, showcasing the number of students receiving each grade (A, B, C, D, F) clearly and comparatively. To uncover deeper relationships, we will focus on the interaction between MBTI type, family income, and academic achievement. Furthermore, histograms were used to analyze the relationship between personality types and their corresponding family income, highlighting trends and correlations among the variables. To further quantify these relationships, the study will generate a correlation coefficient matrix to assist us in comprehending the interplay between MBTI types, family income, and academic performance.   As a matter of fact, the design and statistical analysis methods of this study are intended to offer a solid empirical foundation for understanding the relationship between MBTI type, family income, and academic achievement among undergraduates. This research is anticipated to provide a reference for subsequent related studies and promote more in-depth discussions in the field of education. Through scientific research design and rigorous statistical analysis, we aim to reveal the internal links between these variables and provide theoretical support for optimizing educational policies and practices.

\section{Result}\label{sec:result}
\subsection{Overall MBTI distribution}
\begin{figure}[htbp]
    \centering
    \begin{minipage}[b]{0.49\textwidth}
        \centering
        \includegraphics[width=\textwidth]{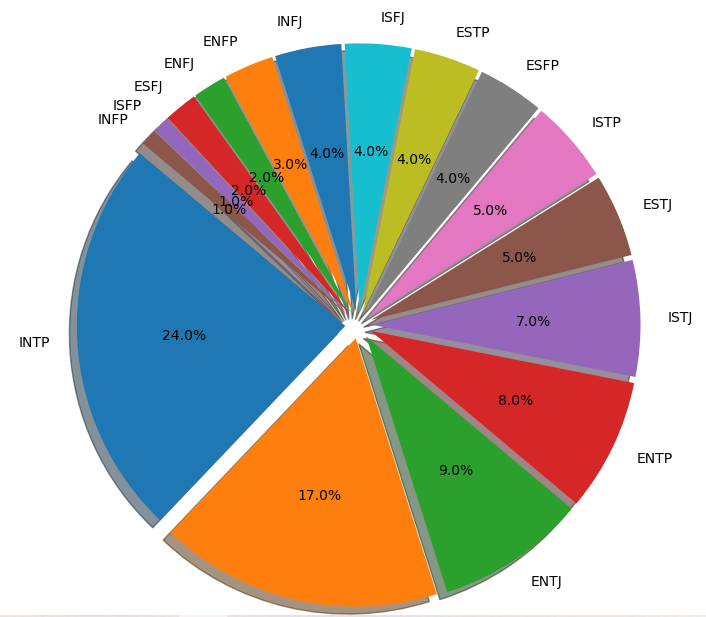}
        \label{fig:sub1}
    \end{minipage}
    \hfill
    \begin{minipage}[b]{0.49\textwidth}
        \centering
        \includegraphics[width=\textwidth]{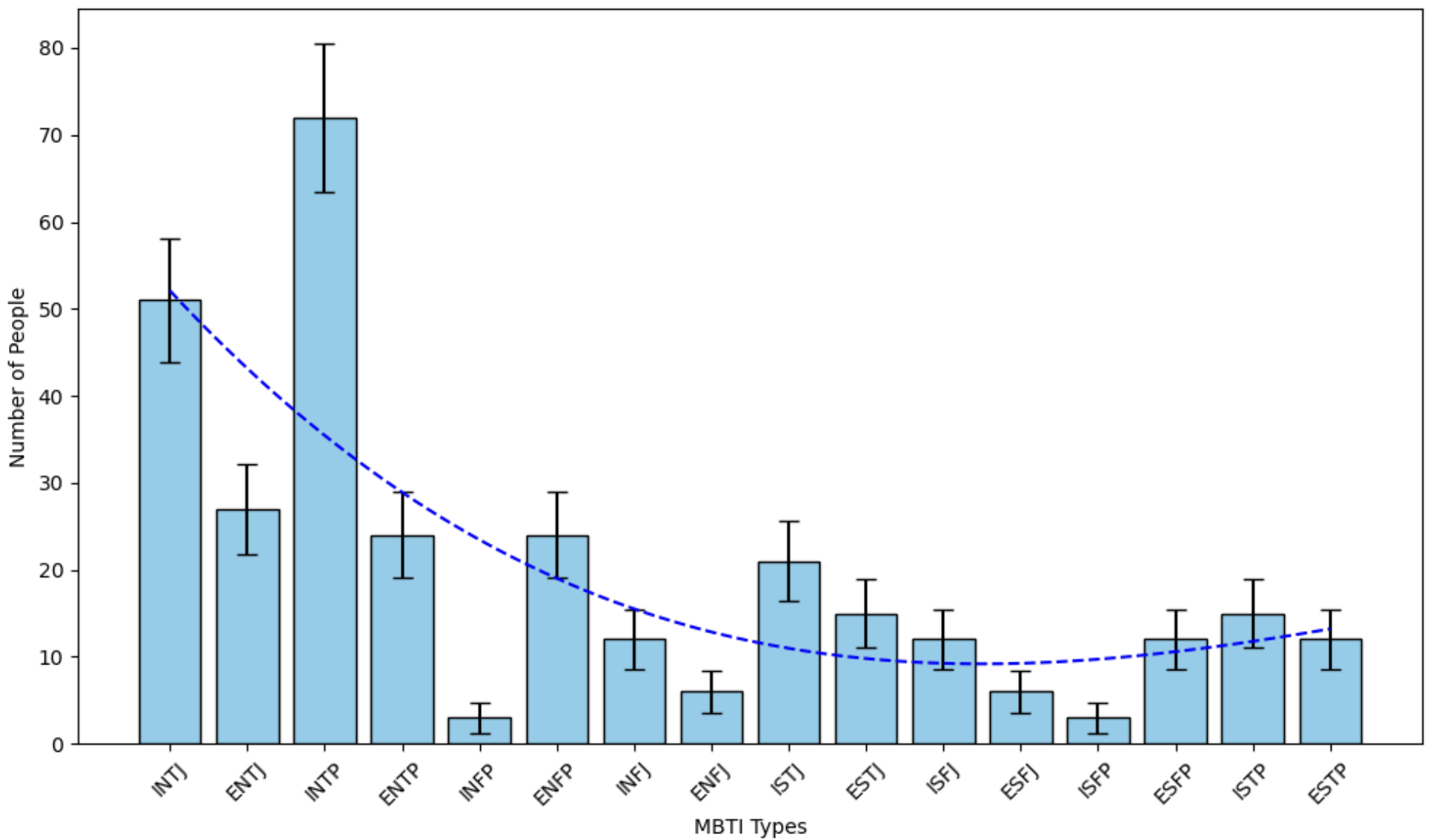}
        \label{fig:sub2}
    \end{minipage}
    \caption{The MBTI distribution graph, a) the pie chart of MBTI personality distribution in Department of Physics, Zhejiang University, b) Histogram with errorbar of personality distribution in Department of Physics, Zhejiang University}
    \label{fig:0}
\end{figure}

In a survey conducted among 300 physics undergraduates at Zhejiang University, the distribution of MBTI personality types revealed distinctive personality characteristics within this academic group. The survey indicated that 24\% of students identified as INTP, a personality type marked by logical thinking, strong analytical skills, and a preference for working independently. This group constituted the largest proportion of respondents. Following closely, 17\% of students were classified as INTJ, known for their strategic thinking, high standards, and focus on long-term goals. Together, INTP and INTJ students accounted for over 40\% of the sample, suggesting a trend toward introversion as well as intellectual and strategic traits among the physics students. Furthermore, 9\% of students were classified as ENTJ, frequently associated with leadership, efficiency, and goal-oriented behavior, while 8\% were identified as ENTP, known for their creativity, adaptability, and passion for exploring new ideas. Additionally, 7\% of students were ISTJ, typically characterized by meticulousness, pragmatism, and high reliability. Other personality types, such as ESTJ (5\%), displayed a tendency toward organization and structured environments. Personality types ESFP, ESTP, ISTP, ISFJ, INFJ, and ENFJ were each represented by 4\% to 5\% of the respondents. In addition, 3\% of students were ENFP, a type recognized for enthusiasm, creativity, and an open mindset, whereas 2\% were categorized as ESFJ, known for sociability and attentiveness to others’ needs. Finally, the rarest personality types within the sample were ISFP and INFP, each comprising only 1\% of the sample. This rarity suggests that students with stronger intrinsic values and introverted, emotionally driven decision-making styles were relatively uncommon among the physics undergraduates. The complete distribution of personality types is depicted in Figure \ref{fig:0}.
\subsection{grades distribution with respect to MBTI}
\begin{figure}[ht]
    \centering
    \begin{minipage}[b]{0.49\textwidth}
        \centering
        \includegraphics[width=\textwidth]{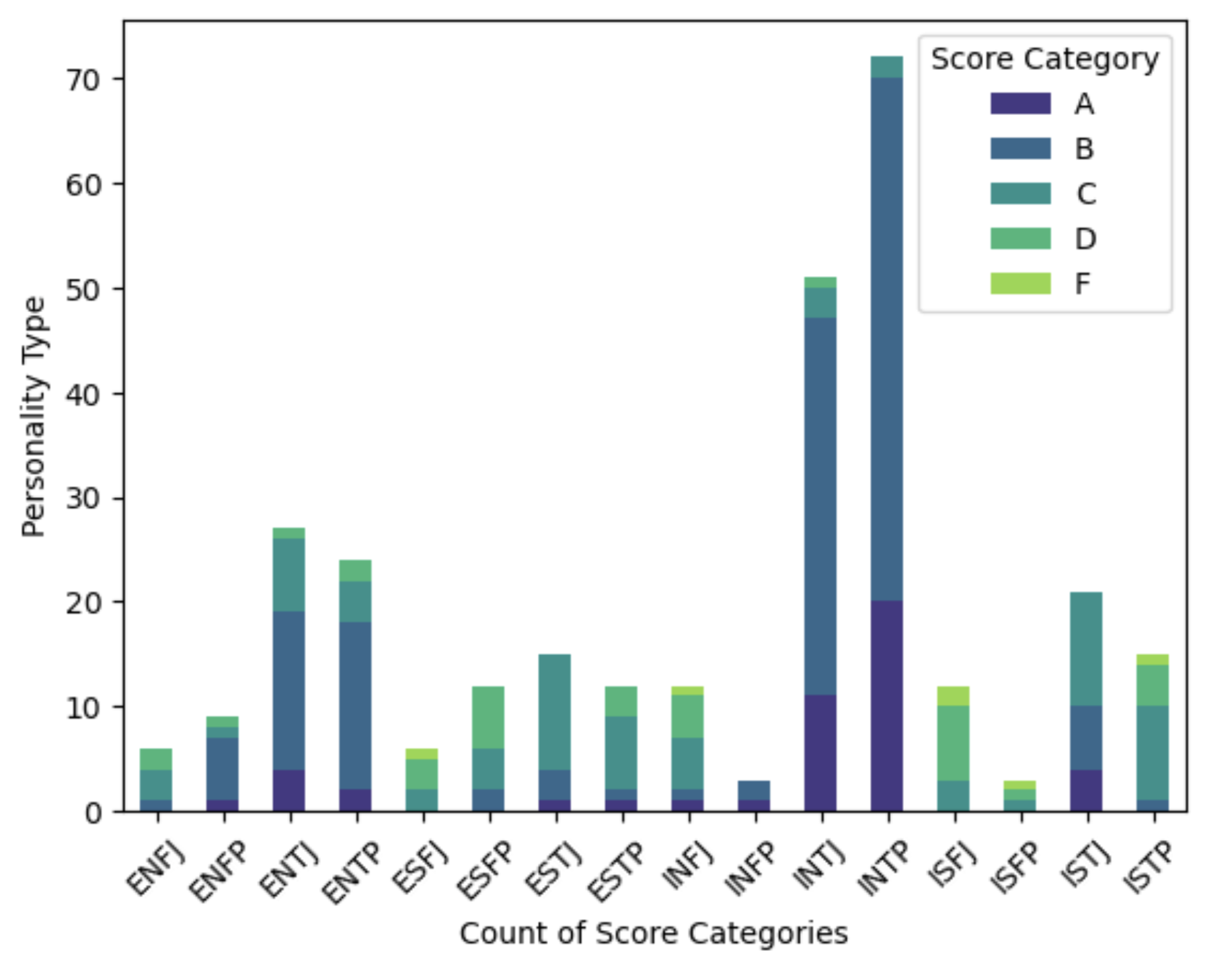}
        \label{fig:sub2}
    \end{minipage}
    \hfill
    \begin{minipage}[b]{0.5\textwidth}
        \centering
        \includegraphics[width=\textwidth]{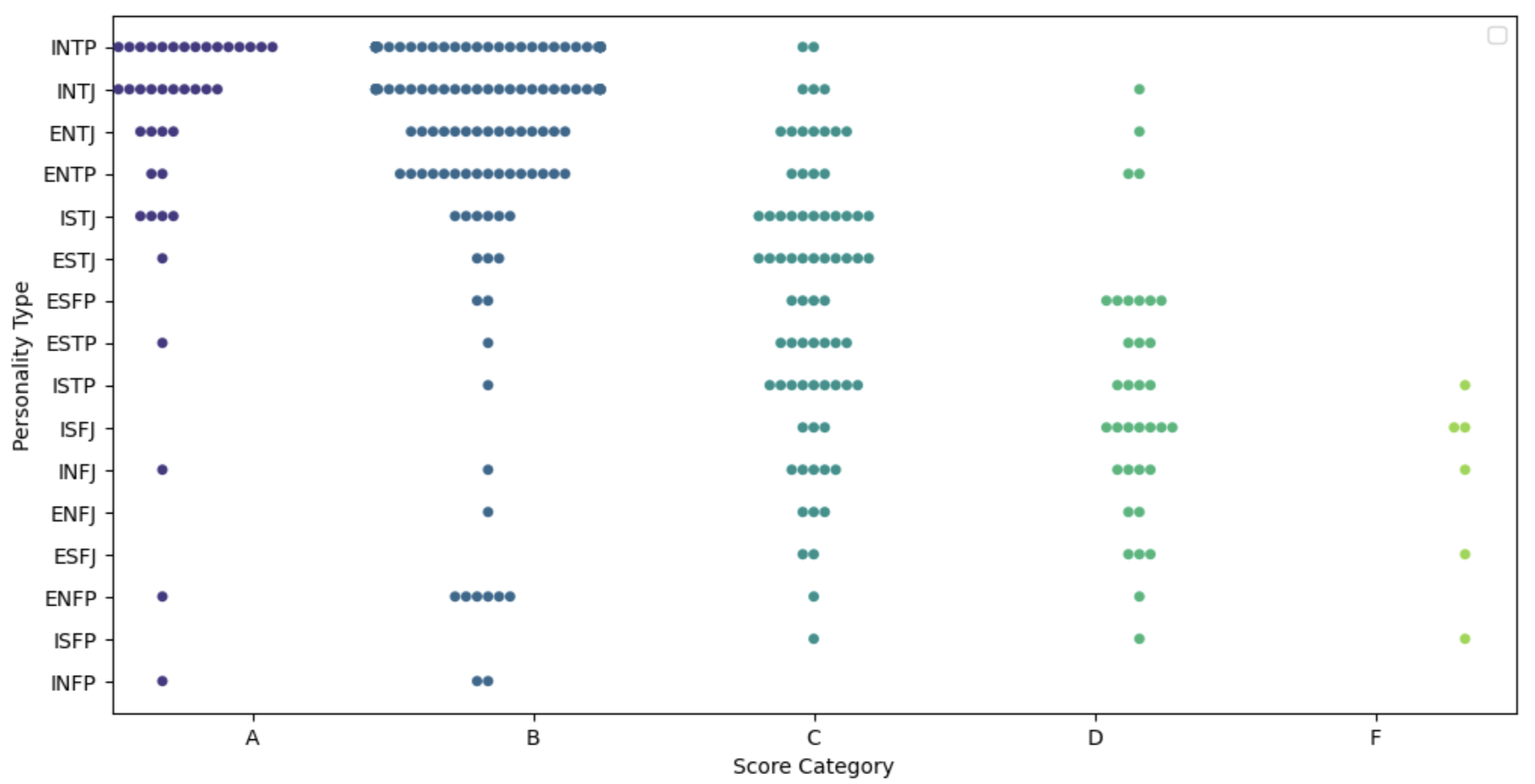}
        \label{fig:sub2b}
    \end{minipage}
    \caption{The MBTI personality grade distribution graph, a) The histogram of MBTI personality grade distribution in Department of Physics, Zhejiang University, b) The scatter plot of MBTI personality grade distribution in Department of Physics, Zhejiang University}
    \label{fig:2}
\end{figure}

An analysis of the academic performance of undergraduate physics students classified by the Myers-Briggs Type Indicator (MBTI) personality type reveals significant disparities in grade distribution (A, B, C, D, F), as depicted in Table \ref{tab:1} and Figure \ref{fig:2}. Students of the INTP personality type achieved the highest grades, with 20 As and 50 Bs, suggesting a strong correlation between this personality type and academic success. This performance can be attributed to the analytical and innovative thinking typically exhibited by INTPs, which is highly compatible with the requirements of physics courses\citep{poropat2009meta}. The INTJ type also demonstrated commendable grades, with 11 As and 36 Bs, indicating that their strategic planning and organizational capabilities contribute to their academic excellence \citep{vesely2014ei}. 
\begin{table}[htbp]
    \centering
    \tabletypesize{\scriptsize} 
    \caption{Score Counts by Personality Type} \label{tab:1}
    \setlength{\tabcolsep}{3pt} 
    \resizebox{\textwidth}{!}{%
    \begin{tabular}{lcccccccccccccccc}
        \hline
        \textbf{Grade} & \textbf{ENFJ} & \textbf{ENFP} & \textbf{ENTJ} & \textbf{ENTP} & \textbf{ESFJ} & \textbf{ESFP} & \textbf{ESTJ} & \textbf{ESTP} & \textbf{INFJ} & \textbf{INFP} & \textbf{INTJ} & \textbf{INTP} & \textbf{ISFJ} & \textbf{ISFP} & \textbf{ISTJ} & \textbf{ISTP} \\
        \hline
        \textbf{A} & 0 & 1 & 4 & 2 & 0 & 0 & 1 & 1 & 1 & 1 & 11 & 20 & 0 & 0 & 4 & 0 \\
        \textbf{B} & 1 & 6 & 15 & 16 & 0 & 2 & 3 & 1 & 1 & 2 & 36 & 50 & 0 & 0 & 6 & 1 \\
        \textbf{C} & 3 & 1 & 7 & 4 & 2 & 4 & 11 & 7 & 5 & 0 & 3 & 2 & 3 & 1 & 11 & 9 \\
        \textbf{D} & 2 & 1 & 1 & 2 & 3 & 6 & 0 & 3 & 4 & 0 & 1 & 0 & 7 & 1 & 0 & 4 \\
        \textbf{F} & 0 & 0 & 0 & 0 & 1 & 0 & 0 & 0 & 1 & 0 & 0 & 0 & 2 & 1 & 0 & 1 \\
        \hline
    \end{tabular}%
    }
\end{table}

Moderate academic performance is characteristic of the ENTP and ENTJ personality types. Specifically, ENTP students achieved 2 A's and 16 B's, while ENTJ students earned 4 A's and 15 B's. These results highlight their strengths in problem-solving and leadership within an academic setting, though they fall short of the exceptional performance observed among INTP and INTJ types \citep{zhang2002thinking}. In contrast, certain personality types, such as ISFJ and ISTP, show more concerning trends. ISFJ students received a total of 7 D's and 2 F's, suggesting that they encounter substantial challenges in academic contexts. This could be attributed to their preference for structured and practical tasks over theoretical exploration \citep{poropat2009meta}. Similarly, ISTP students exhibited a concerning performance, with grades distributed across 9 C's, 4 D's, and 1 F, which may indicate difficulties in adapting to the abstract nature of physics research \citep{myers1998}. Overall, these findings emphasize that intuitive personality types, such as INTP and INTJ, generally perform better academically than sensory types, like ISFJ and ISTP. This pattern aligns with existing literature, which underscores the advantages of intuitive thinking in fields requiring complex problem-solving and innovative approaches \citep{zhang2002thinking}, \citep{myers1998}. Additionally, extroverted and thinking-oriented students, such as those with ENTJ and ENTP types, tend to excel in collaborative academic settings \citep{white1982relation}. These results underscore the importance of understanding personality traits in educational contexts to support diverse student needs through tailored academic resources.

\subsection{distribution of NT, NF, SJ, SP among MBTI}
\begin{figure}[h]
    \centering
    \includegraphics[width=0.5\linewidth]{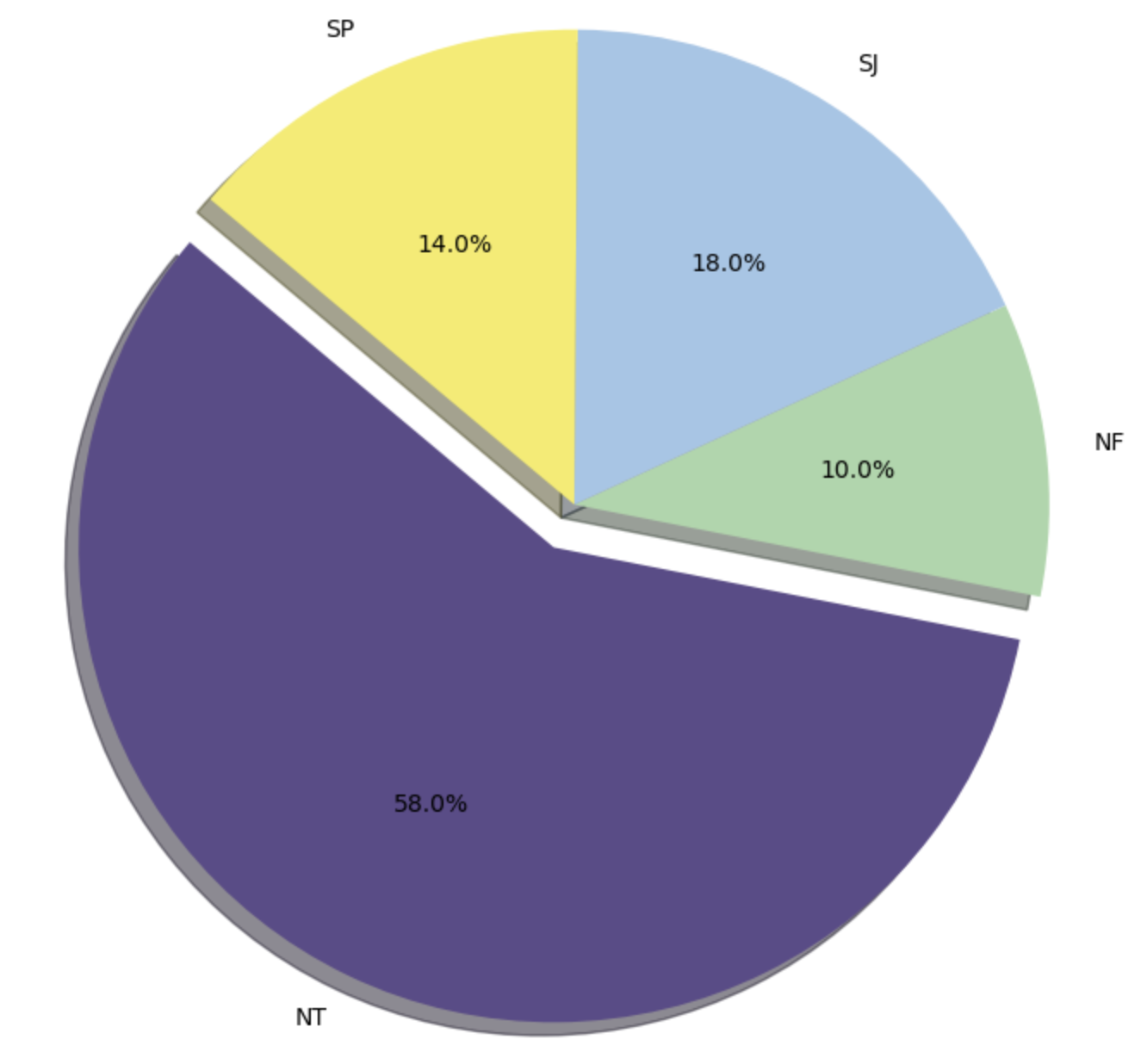}
    \caption{The pie chart of the distribution of NT, NF, SJ, SP among MBTI}
    \label{fig:fig}
\end{figure}

As depicted in Figure \ref{fig:fig}, the analysis of personality categories offers valuable insights into the distribution of personality types within the surveyed population. Notably, the NT (Intuitive Thinking) category, represented by the purple section of the chart, constitutes the largest proportion at 58\%. This suggests that individuals who tend to emphasize logical reasoning, abstract thinking, and strategic planning are prominently present in their decision-making processes. The dominance of the NT category indicates that a considerable proportion of the population might be proficient in analytical tasks and thrive in settings that require innovative problem-solving. 

Aside from the NT (Intuitive-Thinking) type, the SJ (Sensing-Judging) type represents 18\% of the respondents. This group generally exhibits structured and organized behaviors, often favoring established protocols and guidelines in their activities. Their notable presence underscores a preference for order and reliability within this subset of students.  In comparison, the SP (Sensing-Perceiving) category comprises 14\% of respondents and includes more adaptable and spontaneous individuals. Members of this group typically value flexibility and hands-on experiences over rigid structures, contributing to a more dynamic and responsive approach to learning and interaction. Finally, the NF (Intuition-Feeling) category constitutes the smallest segment, accounting for just 10\% of respondents. Characterized by a strong emphasis on emotions and values, this group prioritizes interpersonal relationships and seeks to foster harmonious connections.

\subsection{distribution of willingness of Enter the scientific research field among MBTI}
\begin{figure}[h!]
    \centering
    \includegraphics[width=0.8\linewidth]{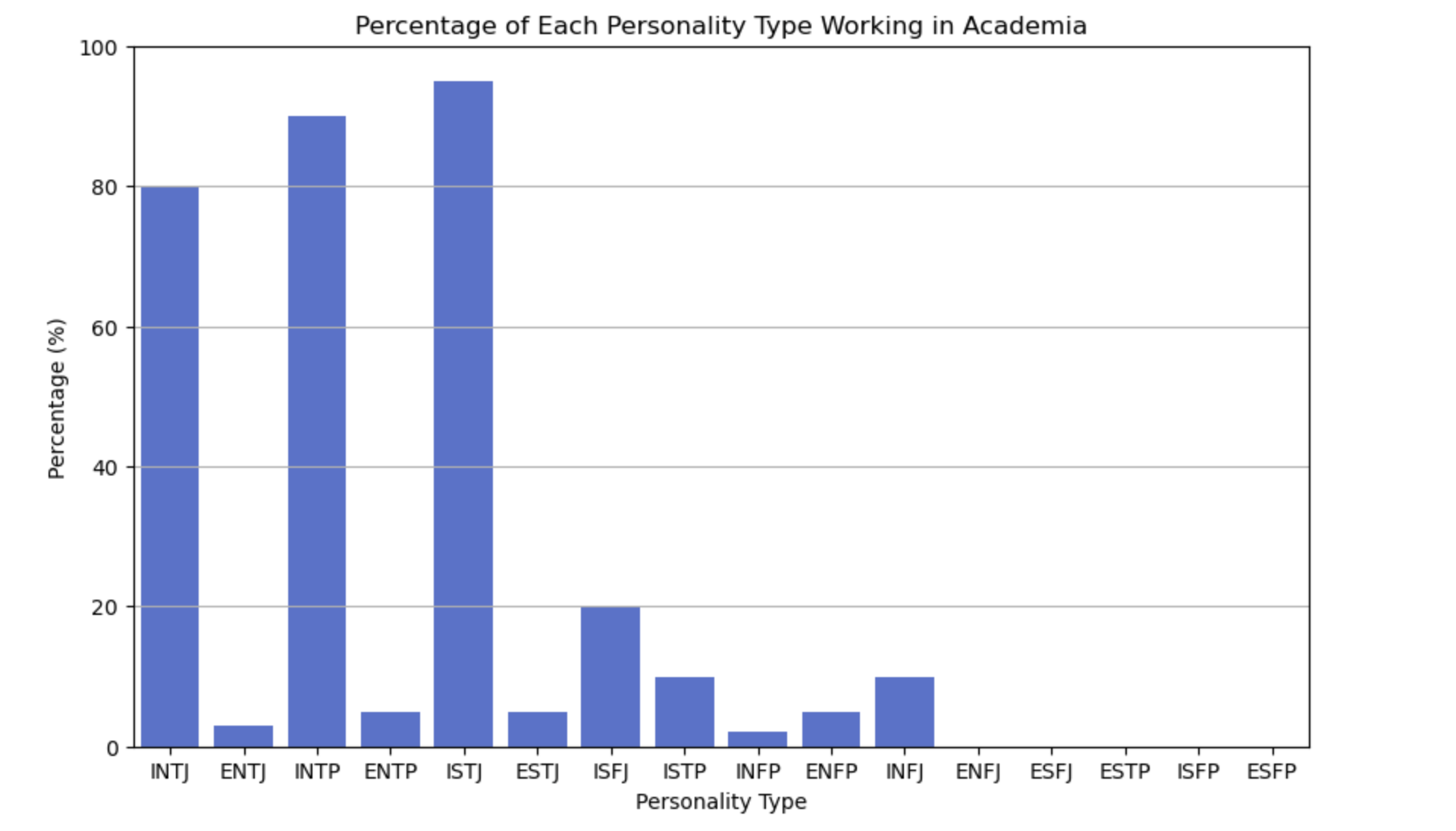}
    \caption{The histogram of willingness of entering the scientific research field among MBTI, x axis represents personality type, y axis represents percentage in this MBTI personality type}
    \label{fig:3}
\end{figure}

There is a notable difference in the willingness to pursue a PhD and academic research across various personality types. As depicted in Figure \ref{fig:3}, the INTJ group demonstrates a particularly strong academic inclination, with 80\% expressing a desire to pursue a doctorate. Similarly, INTP students exhibit high interest, with 90\% indicating a willingness to continue to a doctoral degree. Notably, ISTJ students show the highest level of interest, with 95\% expressing aspirations for further education. In contrast, very few individuals from certain other personality types are inclined toward doctoral studies. Specifically, only 3\% of ENTJ and 5\% of ESTJ, ENTP and ENFP students expressed an interest in pursuing a PhD. Meanwhile, 20\% of ISFJ and 10\% of ISTP and INFJ participants indicated some interest in an academic career. However, students from the remaining personality types, including ESFJ, ESTP, ISFP, and ESFP, showed extremely low interest in pursuing a doctorate, as illustrated in Figure \ref{fig:3}.
These results highlight the correlation between personality types and academic ambitions, supporting previous studies suggesting that personality traits significantly influence educational and career trajectories \citep{mccrae2003}, \citep{mclellan2017personality}. Specifically, the strong inclination of INTP and INTJ individuals toward academia likely reflects their inherent analytical thinking and preference for complex problem-solving---traits essential in research-oriented environments \citep{widiger2013personality}, \citep{o2007big}.

\subsection{Income distribution}
\begin{figure}[h]
    \centering
    \includegraphics[width=0.5\linewidth]{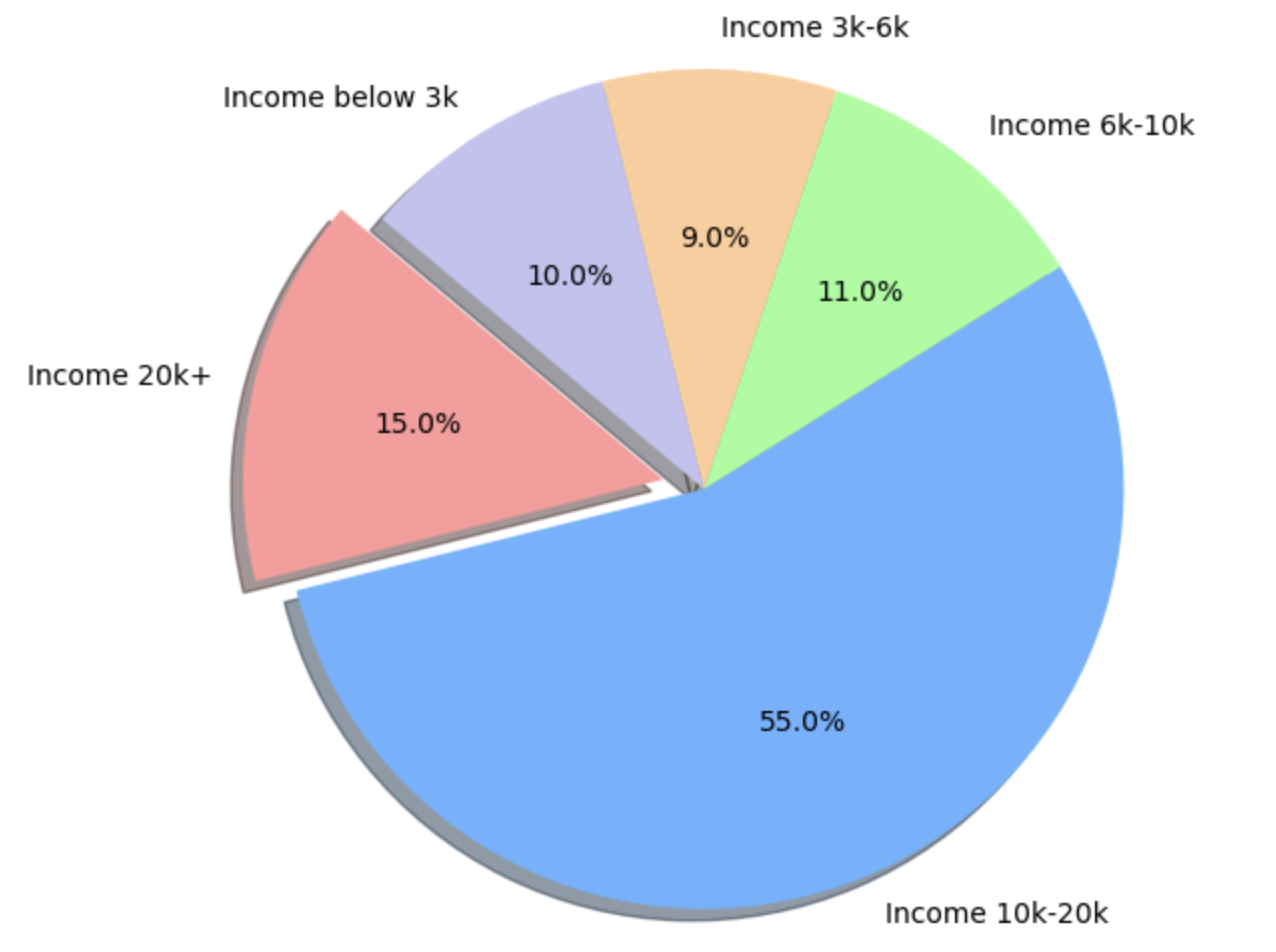}
    \caption{The pie chart of the family monthly income per person for all samples}
    \label{fig:4}
\end{figure}

An analysis of the distribution of parents' income provides valuable insights into the economic backgrounds of the surveyed families. As shown in the pie chart (Figure \ref{fig:4}), the majority of respondents (55\%) reported monthly incomes between ¥10,000 and ¥20,000. This suggests that a significant portion of families are financially stable at a middle level, which may positively impact their children's access to educational opportunities and resources. In contrast, 10\% of families have an income of less than ¥3,000, representing a small segment of low-income households. 
This financial limitation could pose challenges to these families in supporting their children's educational aspirations. Additionally, 9\% of respondents reported incomes between ¥3,000 and ¥6,000, and 11\% reported incomes between ¥6,000 and ¥10,000. These groups represent middle-low income families who may encounter economic constraints that impact their ability to support academic pursuits. Meanwhile, 15\% of households had monthly incomes exceeding ¥20,000, indicating the presence of high-income families within the survey population. This disparity in income distribution highlights the varied economic backgrounds among respondents and suggests potential implications for educational equity. These findings underscore the need for further investigation into the ways parental income affects students' academic performance and career aspirations, particularly for low-income families who may require additional support to navigate the educational path (see Figure \ref{fig:4}).

\section{Discussion}\label{sec:Disscussuion}
We only discuss the five remarkable MBTI personalities, which are INTJ, INTP, ENTJ, ENTP, and ISTJ, since the rest of the MBTI personalities have a sample population of less than 20, which can be neglected.

\begin{figure}[h]
    \centering
    \begin{minipage}[b]{0.49\textwidth}
        \centering
        \includegraphics[width=\textwidth]{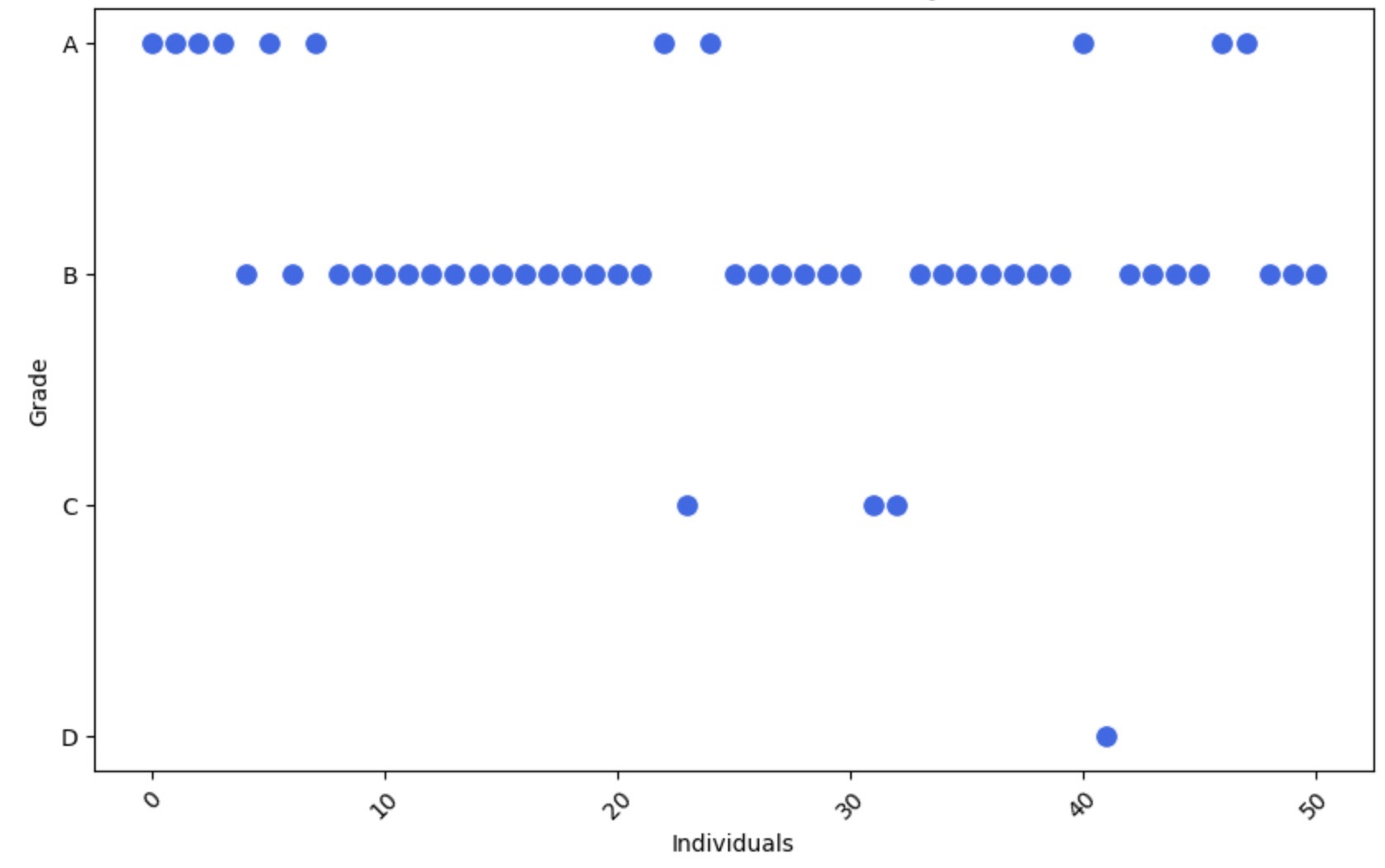}
        \label{fig:sub5}
    \end{minipage}
    \hfill
    \begin{minipage}[b]{0.5\textwidth}
        \centering
        \includegraphics[width=\textwidth]{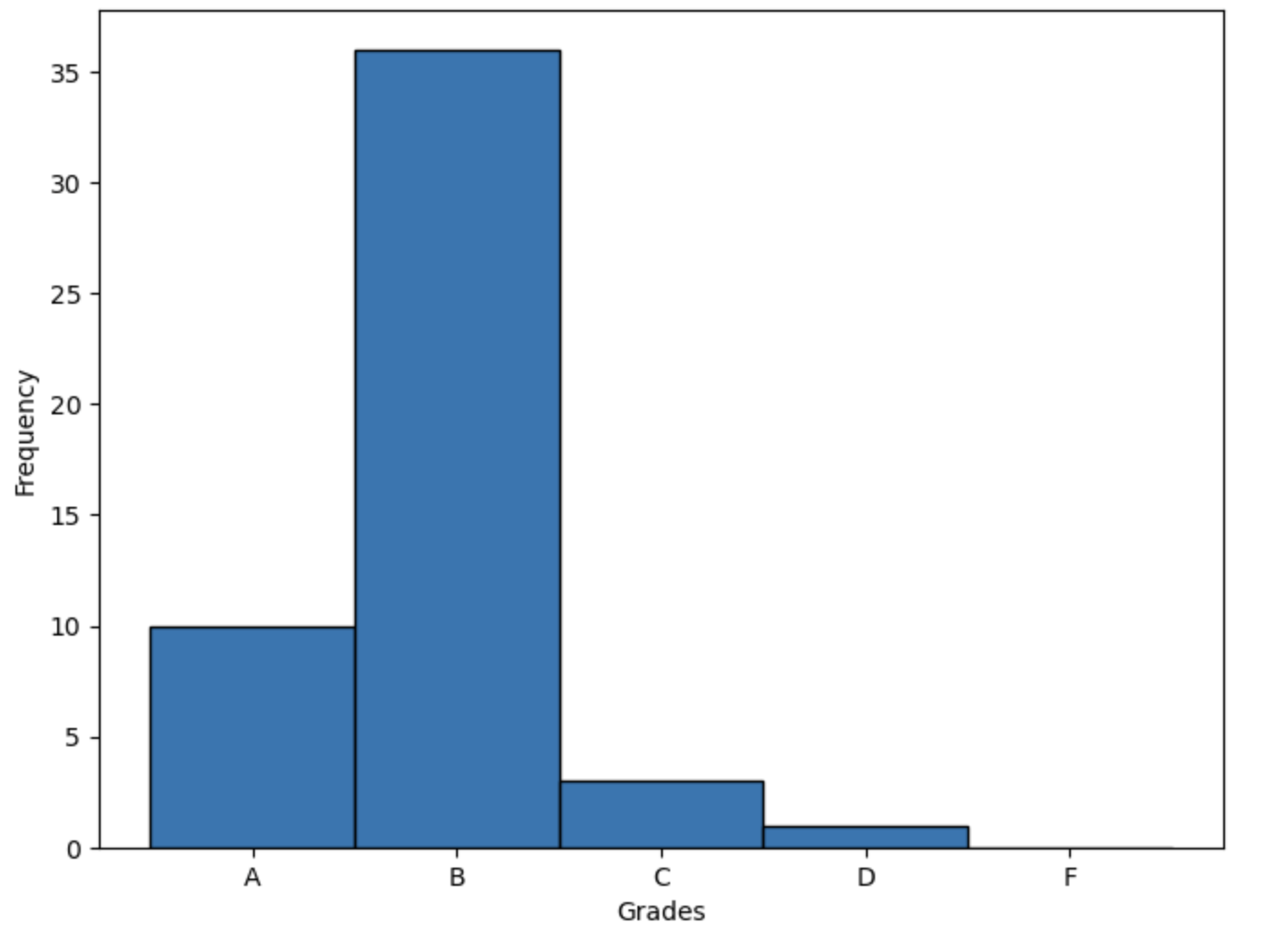}
        \label{fig:sub5b}
    \end{minipage}
    \hfill
    \begin{minipage}[b]{0.5\textwidth}
        \centering
        \includegraphics[width=\textwidth]{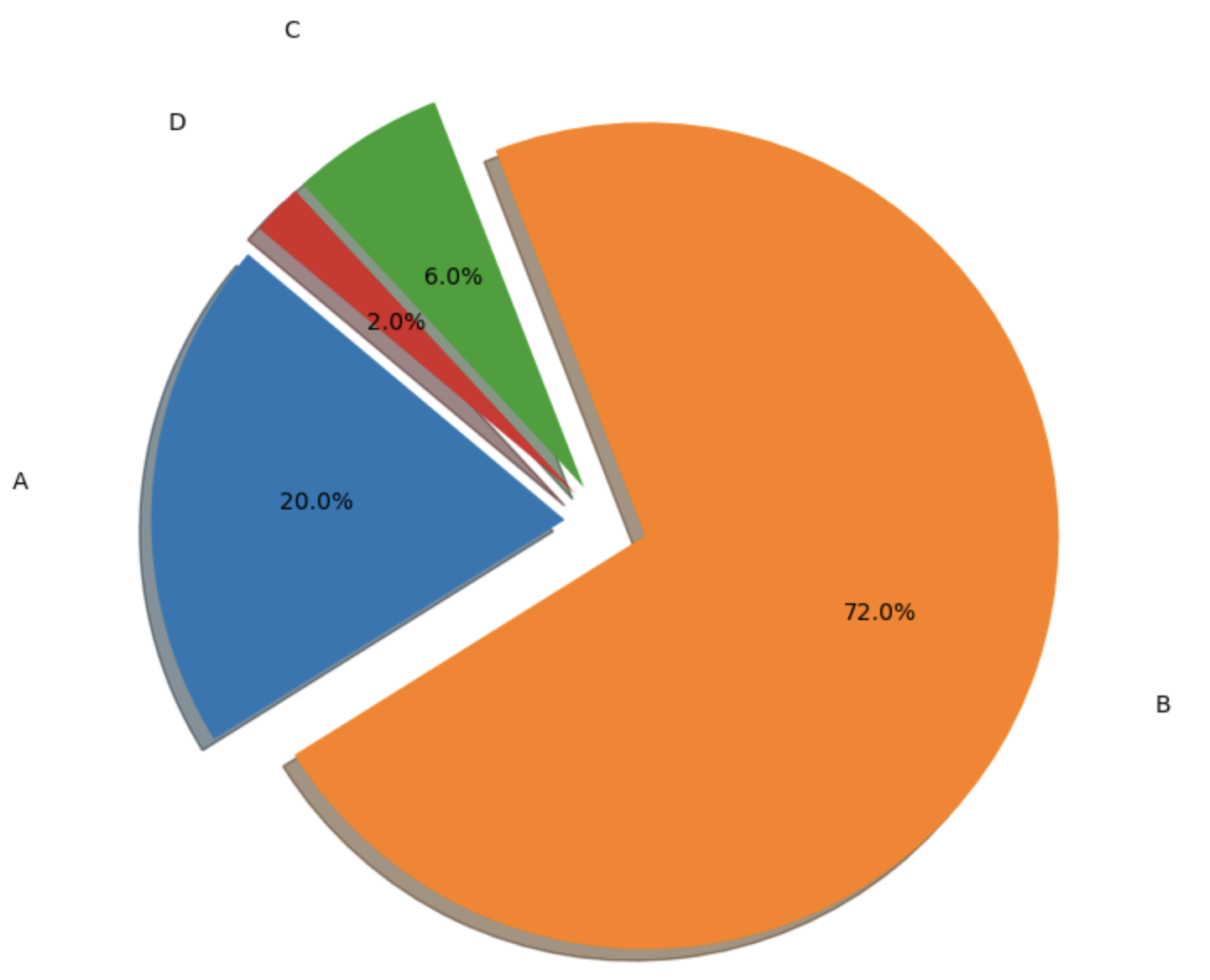}
        \label{fig:sub5c}
    \end{minipage}
    \caption{INTJ grade distribution graph, a) The scatter of INTJ grade distribution in Department of Physics, Zhejiang University, b) The histogram of INTJ personality grade distribution in Department of Physics, Zhejiang University c)The pie chart of INTJ personality grade distribution in Department of Physics, Zhejiang University}
    \label{fig:intj}
\end{figure}
\subsection{Remarkable MBTI personality analysis : INTJ}
First of all, INTJs have impressive academic performance.The majority of INTJ students received a B grade, with 36 individuals (approximately 72.0\%) falling into this category (shown in Figure \ref{fig:intj}). This indicates a strong performance among INTJs, who are often characterized by their strategic thinking, independence, and problem-solving skills \citep{vesely2013teachers}. Notably, 11 individuals (about 20.0\%) earned an A, reflecting a significant portion of high achievers within this personality type. This suggests that INTJs excel academically, leveraging their intuitive and analytical strengths to navigate complex subjects effectively \citep{stein2019evaluating}.

The distribution of lower grades shows a notable decline, with only 3 individuals (around 6\%) receiving a C and a mere 1 individual (approximately 2.0\%) receiving a D. This steep drop-off emphasizes the tendency of INTJs to perform well academically, with very few individuals scoring below a B. The histogram further illustrates that the INTJ personality type, associated with a high degree of independence, critical thinking, and meticulous planning, aligns well with academic success \citep{komarraju2011big}. INTJs possess several characteristics that make them particularly well-suited for academic pursuits. Their ability to think abstractly allows them to engage deeply with theoretical concepts, facilitating a greater understanding of complex material \citep{richardson2012psychological}. Moreover, their strong organizational skills enable them to develop structured study plans and effectively manage their time, essential qualities for navigating the demands of academic life. Additionally, INTJs often exhibit a high level of intrinsic motivation and a commitment to lifelong learning, which drives them to seek out knowledge and explore new ideas within their fields of interest \citep{mccrae2003}.

The limited number of lower grades (C, D and F) could imply that INTJs may struggle in subjects or environments that do not align with their preferred learning styles, particularly those requiring a more flexible and less structured approach. Their preference for systematic analysis and long-term planning may hinder their adaptability in dynamic settings where quick thinking and spontaneity are necessary. Nevertheless, these traits distinctly position INTJs as highly capable individuals in the academic realm, where their analytical abilities and strategic mindset are invaluable assets. Generally speaking, the data indicates that INTJs are likely to adopt effective study strategies and maintain a strong commitment to their academic pursuits. The combination of their cognitive strengths and intrinsic motivation not only enables them to excel academically but also suggests a promising fit for careers in research and academia, where their unique skill set can be fully leveraged to advance knowledge and innovation in their respective fields.

\begin{figure}[h]
    \centering
    \begin{minipage}[b]{0.45\textwidth}
        \centering
        \includegraphics[width=\textwidth]{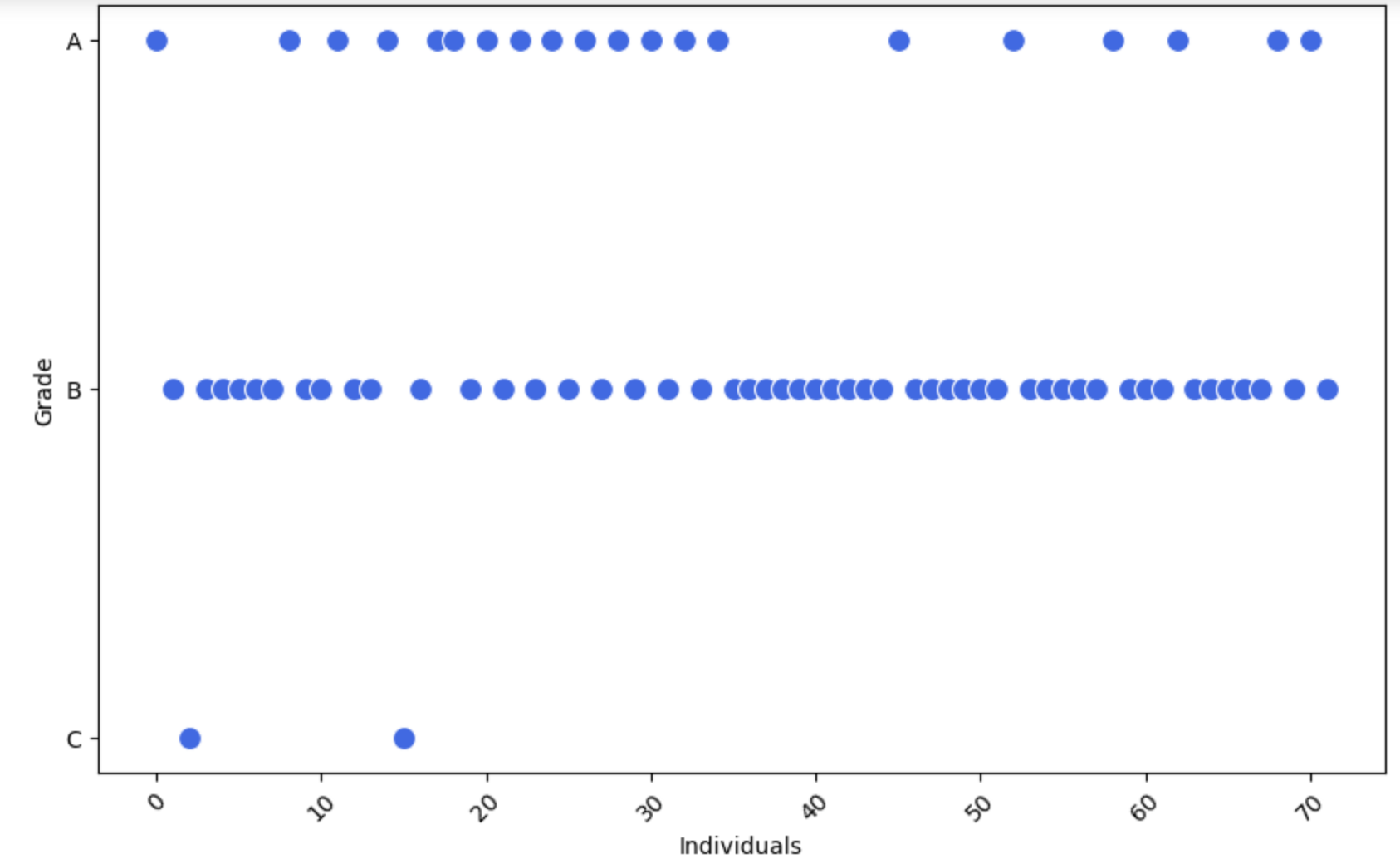}
        \label{fig:sub6}
    \end{minipage}
    \hfill
    \begin{minipage}[b]{0.45\textwidth}
        \centering
        \includegraphics[width=\textwidth]{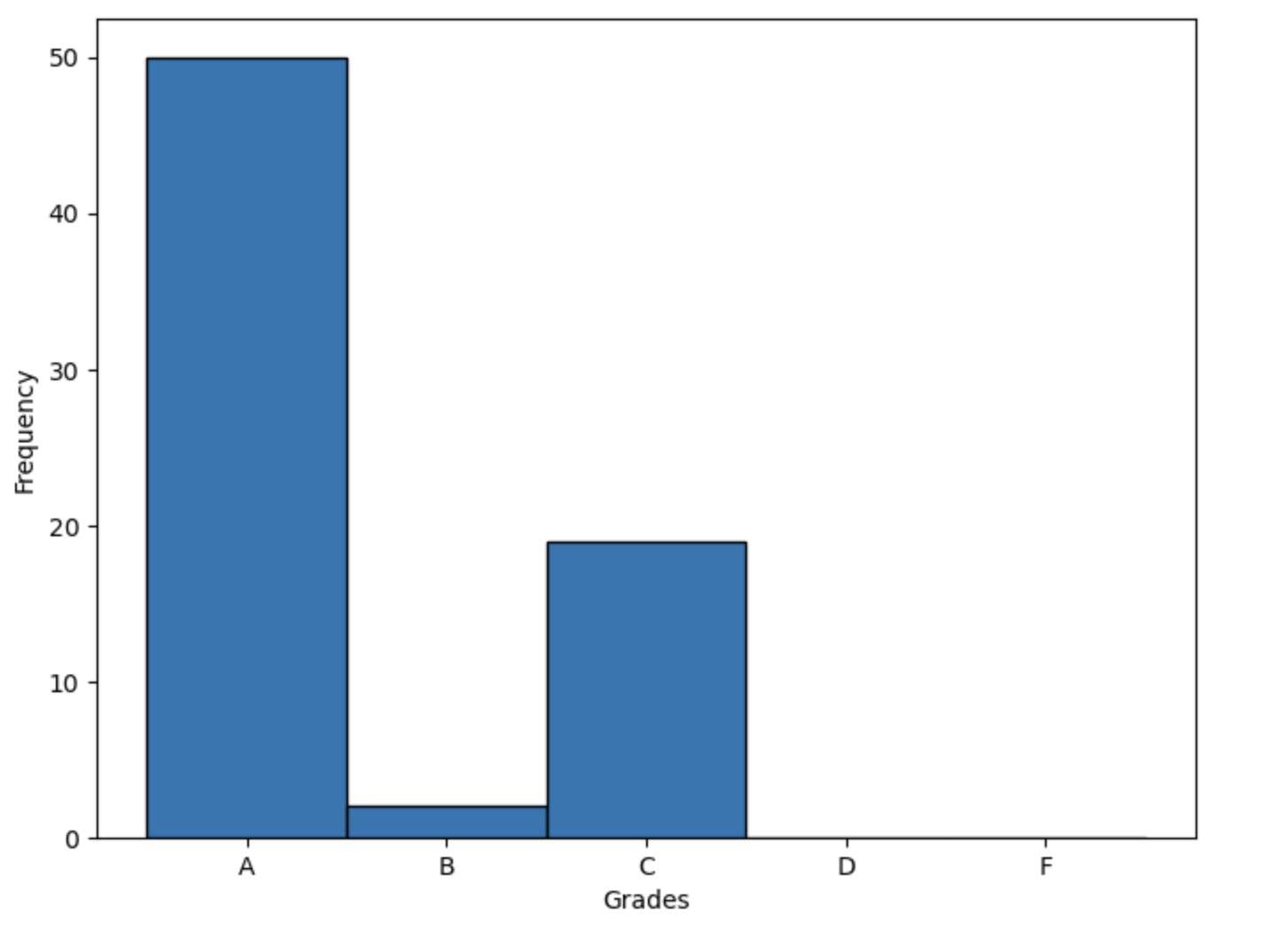}
        \label{fig:sub6b}
    \end{minipage}
    \hfill
    \begin{minipage}[b]{0.45\textwidth}
        \centering
        \includegraphics[width=\textwidth]{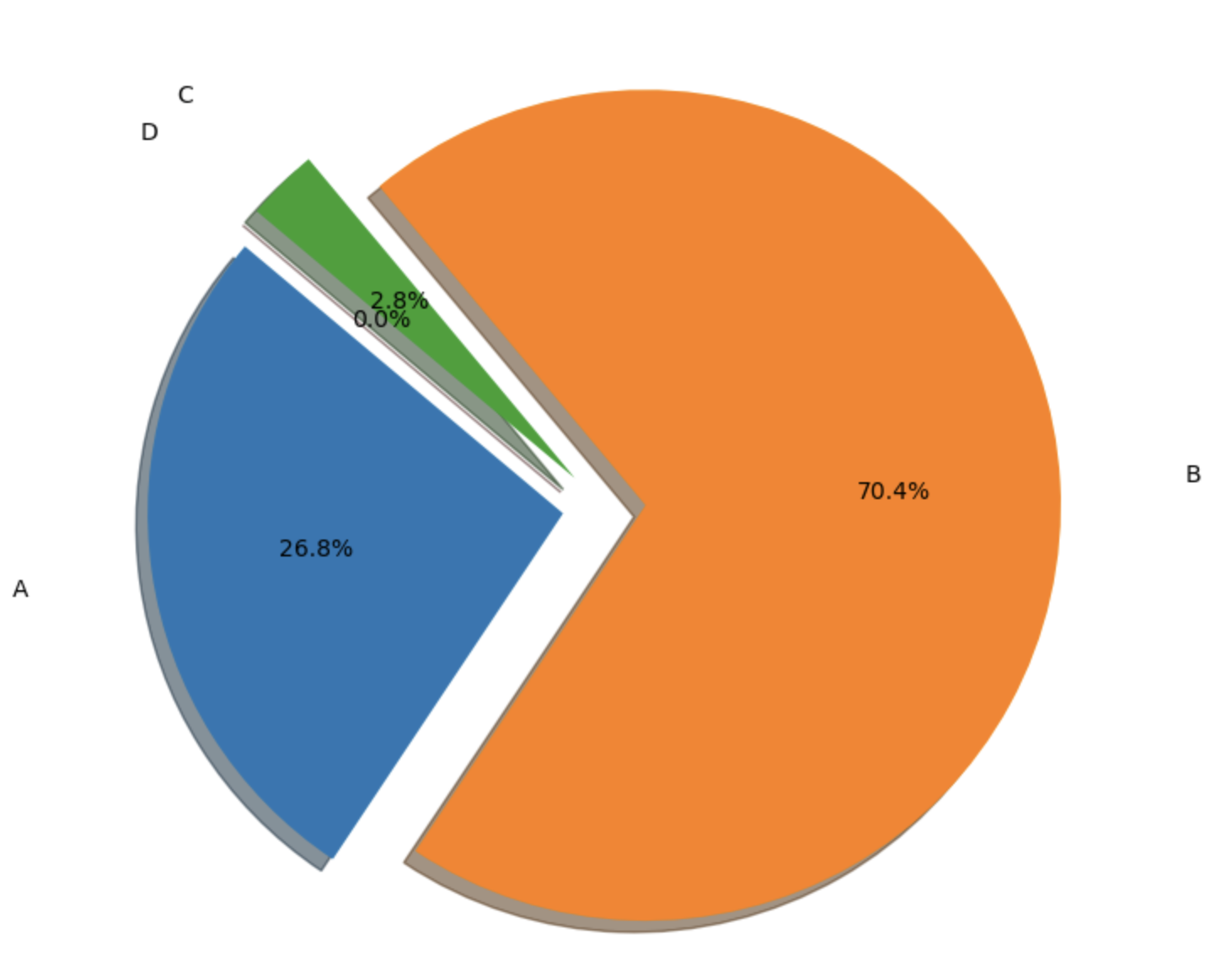}
        \label{fig:sub6c}
    \end{minipage}
    \caption{INTP grade distribution graph, a) The scatter of INTP grade distribution in Department of Physics, Zhejiang University, b) The histogram of INTP personality grade distribution in Department of Physics, Zhejiang University c)The pie chart of INTP personality grade distribution in Department of Physics, Zhejiang University}
    \label{fig:intp}
\end{figure}
\subsection{Remarkable MBTI personality analysis : INTP}
Secondly, INTPs have outstanding academic performance.The analysis of academic performance among INTP students reveals a strong trend toward high achievement, particularly in the B grade category, which has the highest representation, with 50 individuals (approximately 70.4\%) achieving this score (shown in Figure \ref{fig:intp}).  This substantial proportion indicates that INTP students generally perform well academically, reflecting their analytical skills and capacity for independent learning.  These students are known for their love of theory and ability to approach complex problems from multiple angles, traits essential for success in academic settings \citep{furnham2004personality}. In addition, the A grade category includes 20 individuals (around 26.8\%), highlighting a significant, albeit smaller, proportion of high achievers within this personality type.  A grades reinforce the notion that many INTP students excel in their studies, effectively capitalizing on their strengths in problem-solving and abstract thinking.  Their natural curiosity and intellectual rigour drive them to explore concepts deeply, often leading to innovative insights and solutions \citep{matthews2003personality}.

The distribution of lower grades is notably minimal, with only two individuals (about 2.8\%) receiving a C grade.  This suggests that very few INTP students fall below the B grade threshold, indicating a general trend of academic success among this group.  The scarcity of lower grades further emphasizes the alignment of INTP characteristics—such as their preference for theoretical concepts and logical reasoning—with the demands of academic environments that value critical thinking and intellectual engagement \citep{stein2019evaluating}. The scatter plot illustrates this dominance of B grades, suggesting that INTP students maintain a consistently strong performance level in their academic pursuits.  The few C grades indicate that while some students may face challenges, the vast majority of INTPs thrive in environments that encourage independent inquiry and innovative problem-solving.  Their ability to think abstractly and apply logical reasoning allows them to excel in complex subjects, making them particularly well-suited for careers in academia and research \citep{smidt2015big}.

Generally speaking, the combination of analytical thinking, intellectual curiosity, and a propensity for independent learning positions INTP students as strong candidates for academic success.  These traits not only enhance their ability to navigate the challenges of higher education but also foster a lifelong commitment to learning and exploration, essential qualities for thriving in the academic realm.

\begin{figure}[ht]
    \centering
    \begin{minipage}[b]{0.45\textwidth}
        \centering
        \includegraphics[width=\textwidth]{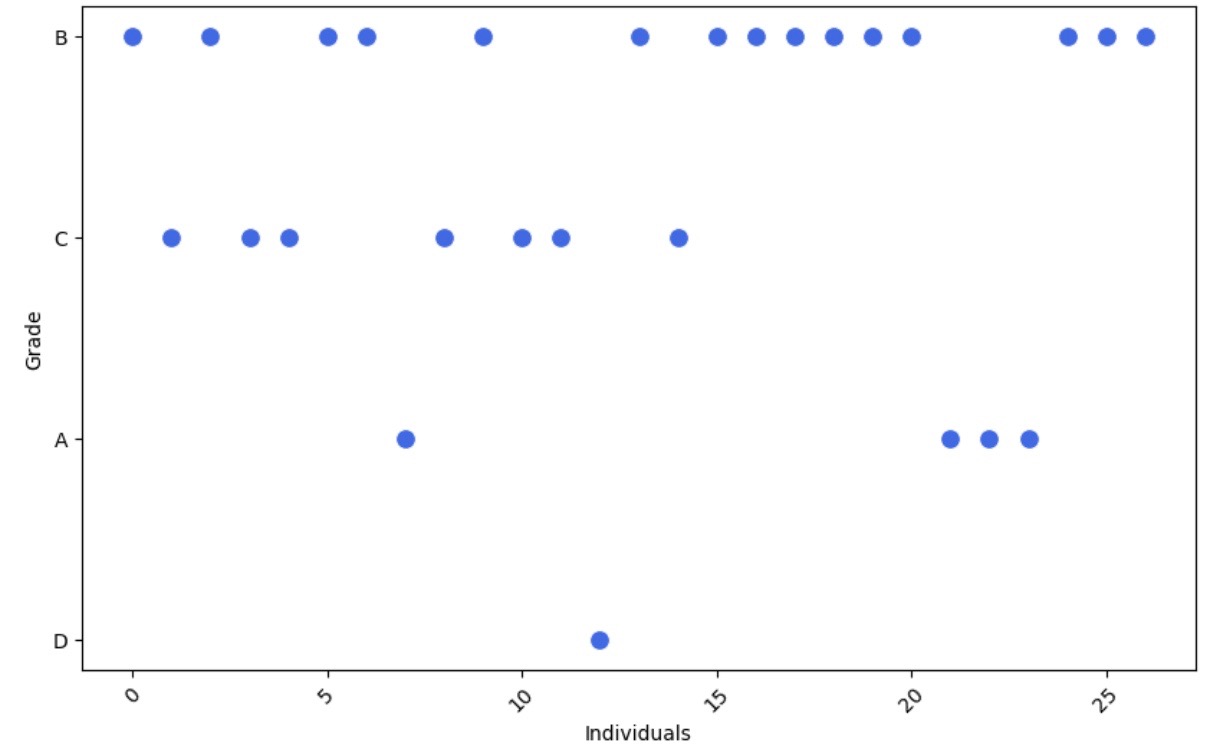}
        \label{fig:sub7}
    \end{minipage}
    \hfill
    \begin{minipage}[b]{0.45\textwidth}
        \centering
        \includegraphics[width=\textwidth]{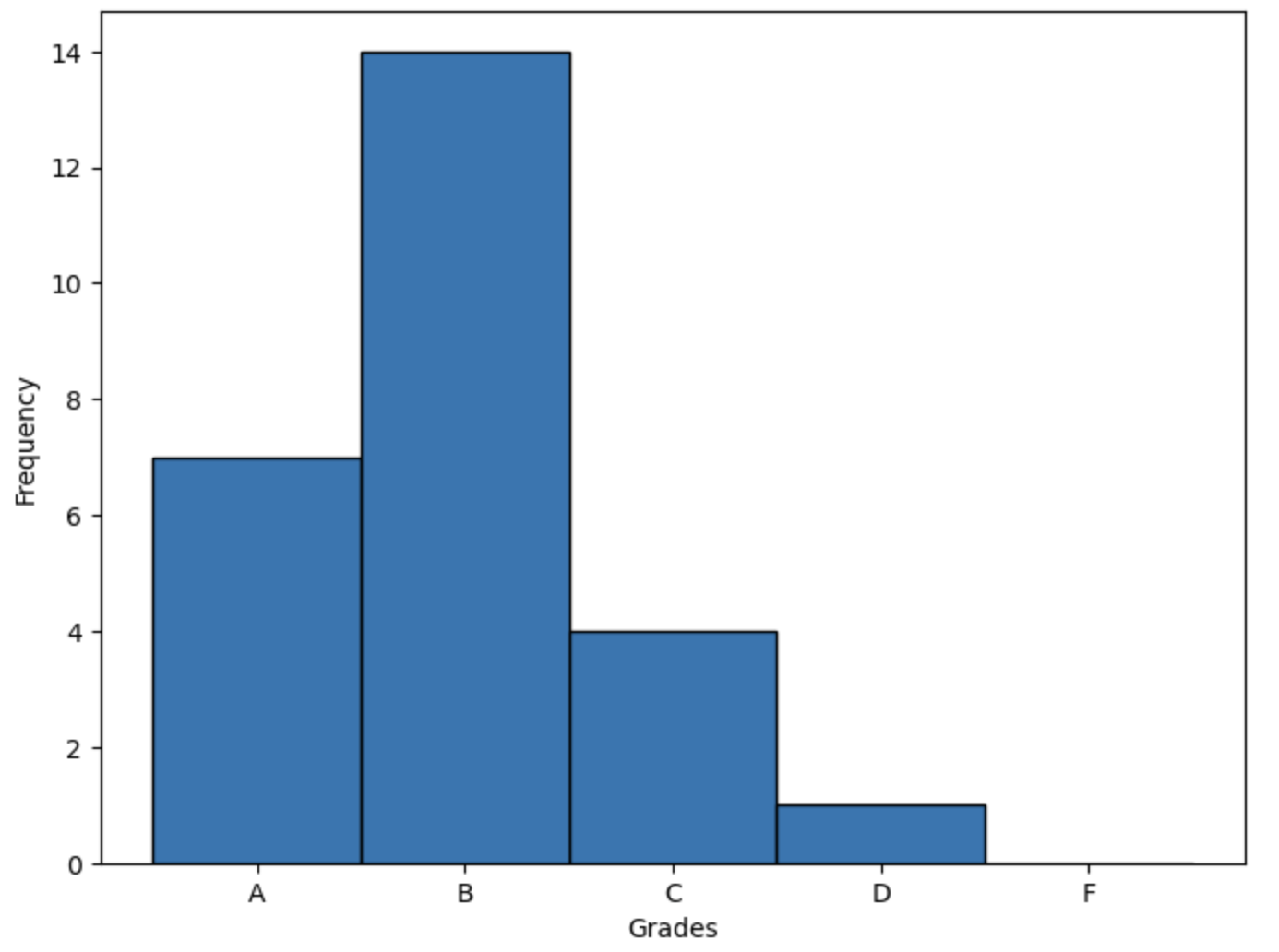}
        \label{fig:sub7b}
    \end{minipage}
    \hfill
    \begin{minipage}[b]{0.45\textwidth}
        \centering
        \includegraphics[width=\textwidth]{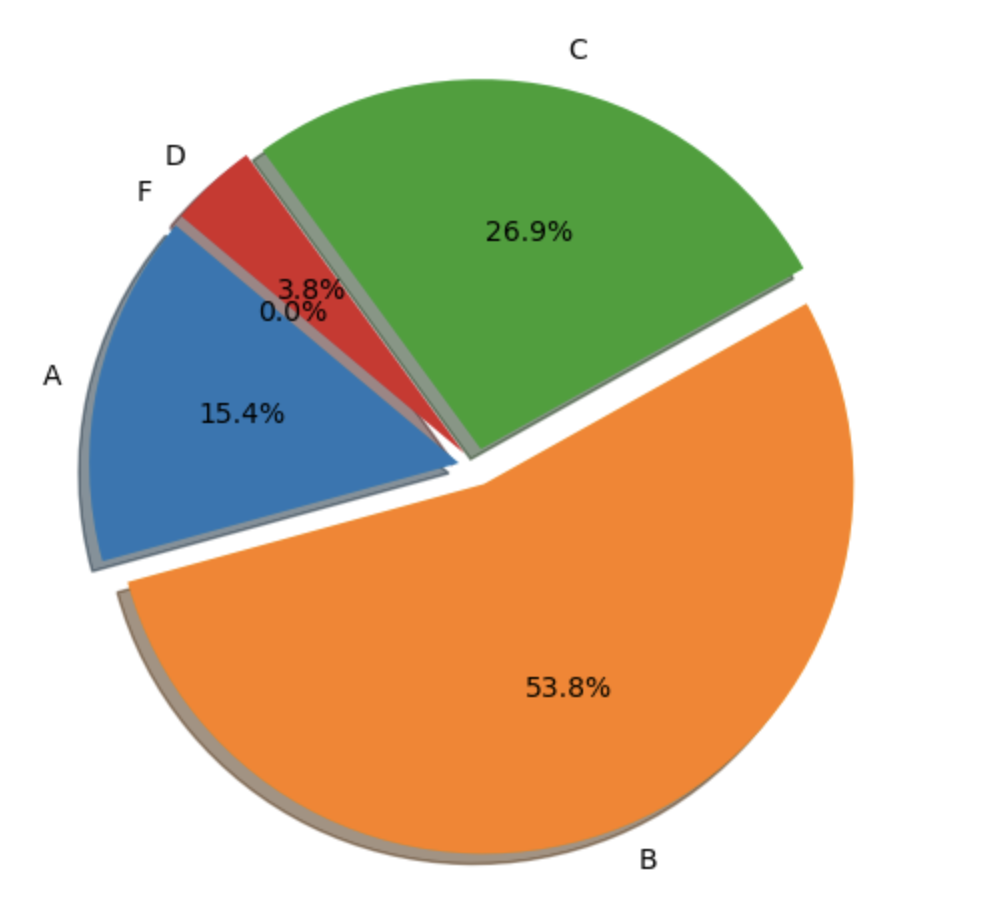}
        \label{fig:sub7c}
    \end{minipage}
    \caption{ENTJ grade distribution graph, a) The scatter of ENTJ grade distribution in Department of Physics, Zhejiang University, b) The histogram of ENTJ personality grade distribution in Department of Physics, Zhejiang University c)The pie chart of ENTJ personality grade distribution in Department of Physics, Zhejiang University}
    \label{fig:entj}
\end{figure}
\subsection{Remarkable MBTI personality analysis : ENTJ}
When it comes to the analysis of ENTJs. The B grade category has a total of 15 individuals (approximately 53.8\%) achieving this score, indicating a moderate level of academic success among ENTJ students (Shown in Figure \ref{fig:entj}).  This suggests that they generally perform well in their academic endeavours, as ENTJs are often characterized by their leadership qualities, strong organizational skills, and goal-oriented mindset \citep{heimlich1990measuring}.  However, the C grade category includes 7 individuals (about 26.9\%), reflecting a significant number of students who may be facing challenges in particular subjects or assessments, indicating room for improvement.  Moreover, 4 individuals (approximately 15.4\%) received an A, which is relatively low compared to the number of B and C grades.  This suggests that while some ENTJs excel, the majority still need to reach the highest academic echelon in this sample.  Notably, there is only 1 individual in the D grade category, indicating that very few ENTJ students performed at the lowest academic level within this group.

The data indicates that ENTJs, while exhibiting some strong performances with B grades, face more variability in their academic outcomes than other personality types. The presence of C grades may suggest potential difficulties arising from their preference for structured and organized approaches, which might not align with all academic challenges. However, the scarcity of A grades among ENTJs could imply that these students may benefit significantly from more supportive educational environments tailored to their leadership and strategic strengths.  This underscores the importance of tailored support in helping ENTJ students reach their full academic potential. Despite the challenges faced by some ENTJs, their traits, including decisiveness and the ability to plan strategically, can make them well-suited for academic careers, especially in leadership roles \citep{stein2019evaluating}.  Research suggests that ENTJs often excel in environments where they can take initiative and drive projects forward, skills that are critical in academic research and organizational settings \citep{chamorro2003personality}.  However, the mixed performance indicated by the presence of C grades suggests that additional support systems are crucial to enhance their academic experiences. These systems can help ENTJ students navigate subjects that do not align with their structured learning preferences, thereby improving their overall academic performance.

While ENTJ students generally show moderate academic success, the variability in their performance highlights the importance of creating educational environments that cater to their strengths while addressing their challenges. By providing tailored support and opportunities for leadership and collaboration, educational institutions can better facilitate the academic journeys of ENTJ students, ultimately enhancing their potential for academic success.

\begin{figure}[h]
    \centering
    \begin{minipage}[b]{0.45\textwidth}
        \centering
        \includegraphics[width=\textwidth]{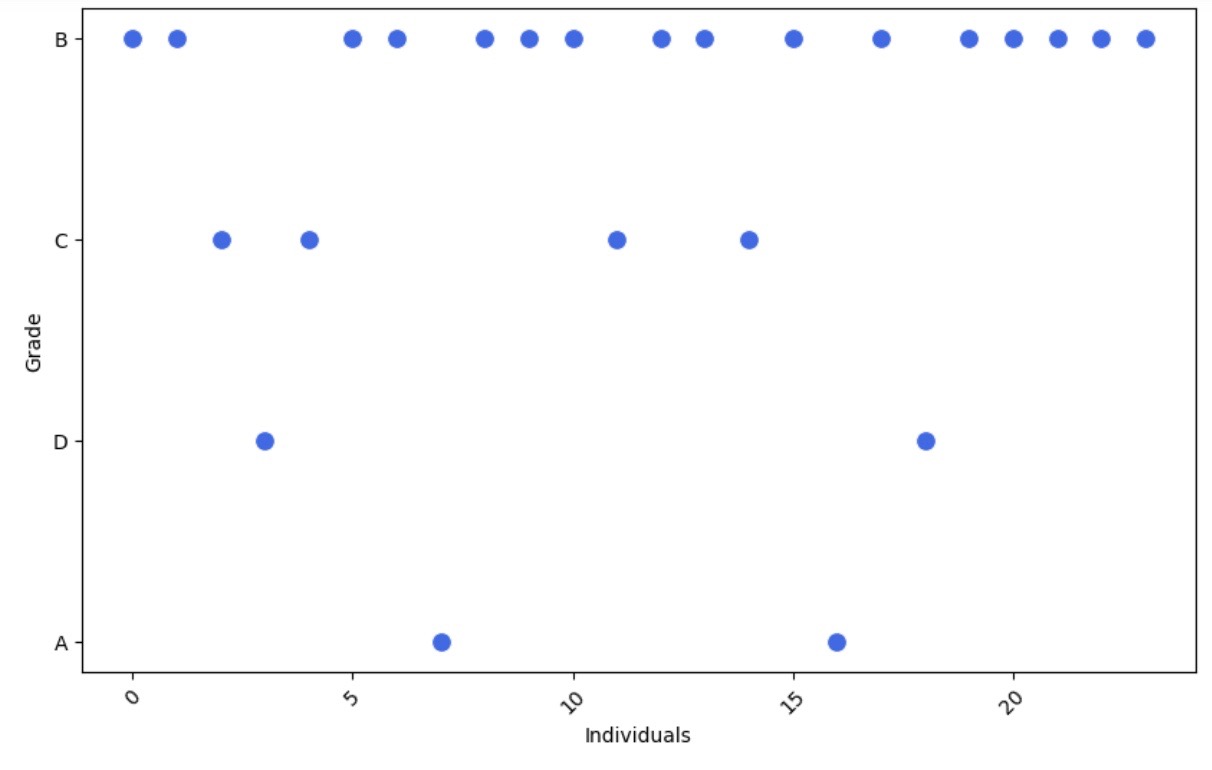}
        \label{fig:sub8}
    \end{minipage}
    \hfill
    \begin{minipage}[b]{0.45\textwidth}
        \centering
        \includegraphics[width=\textwidth]{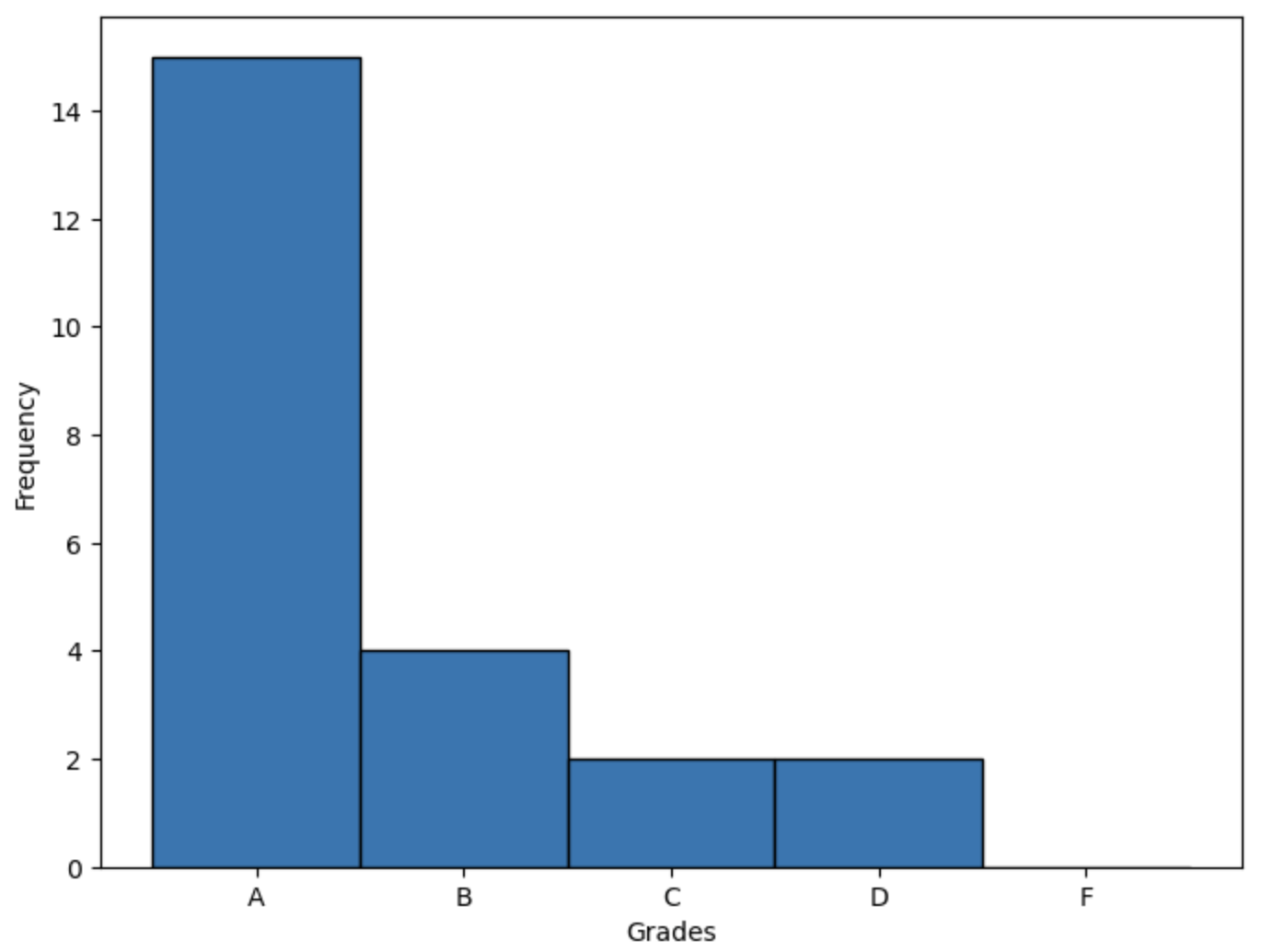}
        \label{fig:sub8b}
    \end{minipage}
    \hfill
    \begin{minipage}[b]{0.45\textwidth}
        \centering
        \includegraphics[width=\textwidth]{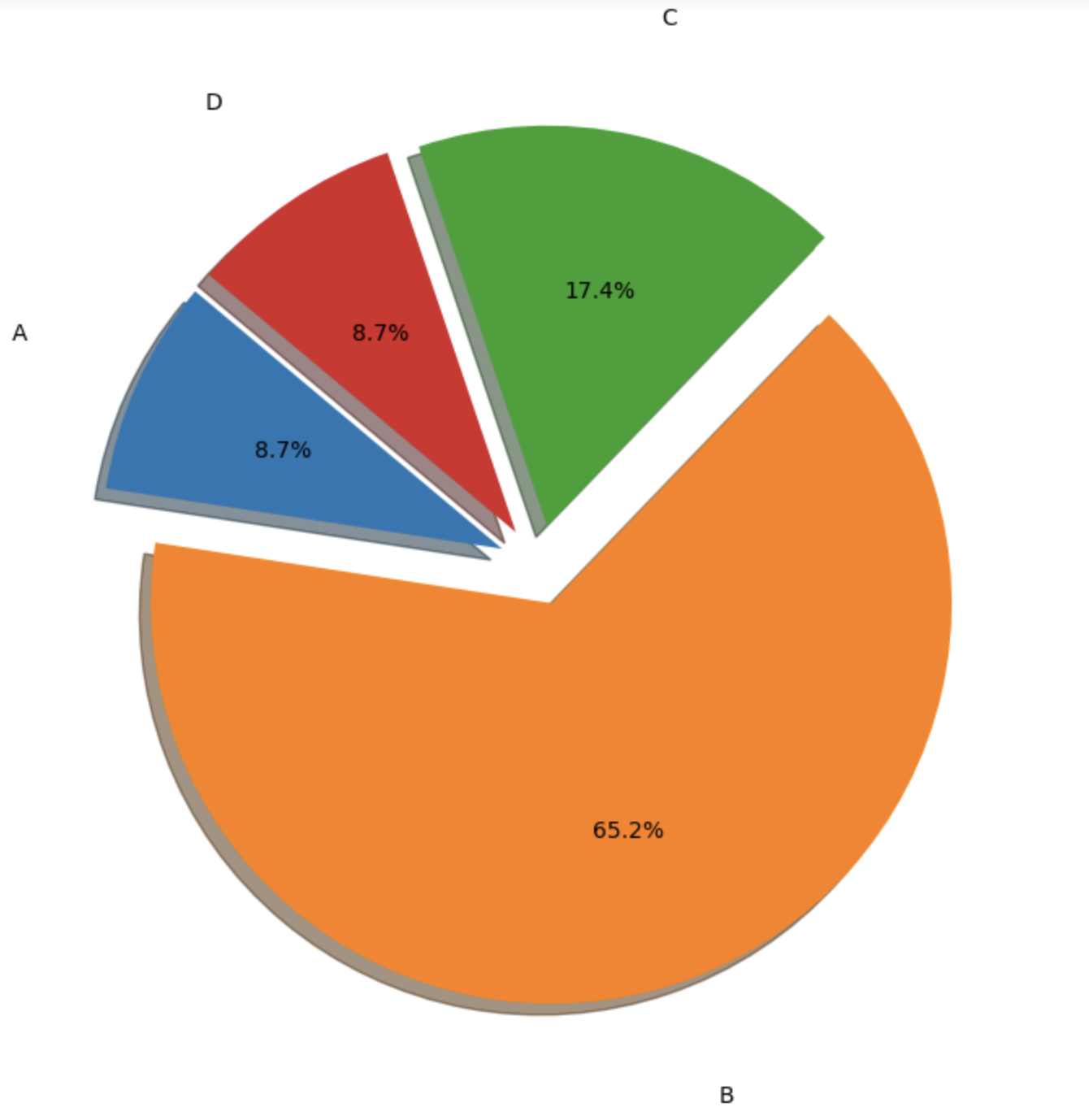}
        \label{fig:sub8c}
    \end{minipage}
    \caption{ENTP grade distribution graph, a) The scatter of ENTP grade distribution in Department of Physics, Zhejiang University, b) The histogram of ENTP personality grade distribution in Department of Physics, Zhejiang University c)The pie chart of ENTP personality grade distribution in Department of Physics, Zhejiang University}
    \label{fig:entp}
\end{figure}
\subsection{Remarkable MBTI personality analysis : ENTP}
ENTPs are an interesting personality to discuss. The analysis of academic performance among ENTP students reveals a notable trend in their grading distribution. The B grade category shows the highest representation, with 16 individuals (approximately 65.2\%) achieving this score (Shown in Figure \ref{fig:entp}). This indicates that a significant portion of ENTP students perform well academically, reflecting their strengths in innovative thinking, adaptability, and enthusiasm for learning. These traits are crucial in academic environments, where the ability to engage with complex concepts and think creatively can lead to successful outcomes \citep{vacha1998reliability}.

The C grade category includes 4 individuals (about 17.4\%), suggesting that some ENTP students may encounter challenges in specific subjects or assessments, which could stem from their preference for novelty and exploration over routine tasks. The A grade category comprises 2 individuals (approximately 8.7\%), indicating a smaller group of high achievers within the ENTP population. This suggests that while some students excel, the majority perform at the B level, consistent with their capability for solid academic performance. Notably, only 2 individuals are in the D grade category, implying that ENTP students rarely perform at the lowest academic level. The predominance of B grades suggests that ENTP students generally maintain a solid performance level, which aligns with their characteristics of being energetic, resourceful, and capable of thinking outside the box\citep{stein2019evaluating}. These qualities enhance their academic experience and position them well for careers in academia, where creativity and innovative problem-solving are valued. The presence of C grades highlights that while many ENTPs are successful, some may struggle in more structured or routine-oriented subjects that require sustained focus and adherence to established guidelines. Their tendency to prefer flexibility and dynamic environments can sometimes conflict with the demands of traditional academic settings, which may necessitate a more systematic approach.

ENTPs thrive in academic environments that encourage exploration and creativity, as these settings allow them to utilize their natural strengths. Their ability to generate novel ideas and engage in critical thinking makes them particularly suited for research-oriented roles, where innovative solutions to complex problems are essential \citep{butterfuss2018role}. However, to fully harness their potential, it's crucial for educational institutions to provide additional support tailored to the unique learning styles of ENTP students. This commitment to understanding and accommodating their needs will help them navigate subjects that require more structure, ensuring their academic success. While ENTP students demonstrate solid academic performance, their distinctive traits also suggest areas where they may encounter challenges. By fostering an educational environment that aligns with their innovative and adaptive nature, institutions can enhance the academic success of ENTP students. This preparation will not only equip them for impactful careers in academia and research but also inspire optimism about the unique contributions they can make in these fields.

\begin{figure}[h]
    \centering
    \begin{minipage}[b]{0.45\textwidth}
        \centering
        \includegraphics[width=\textwidth]{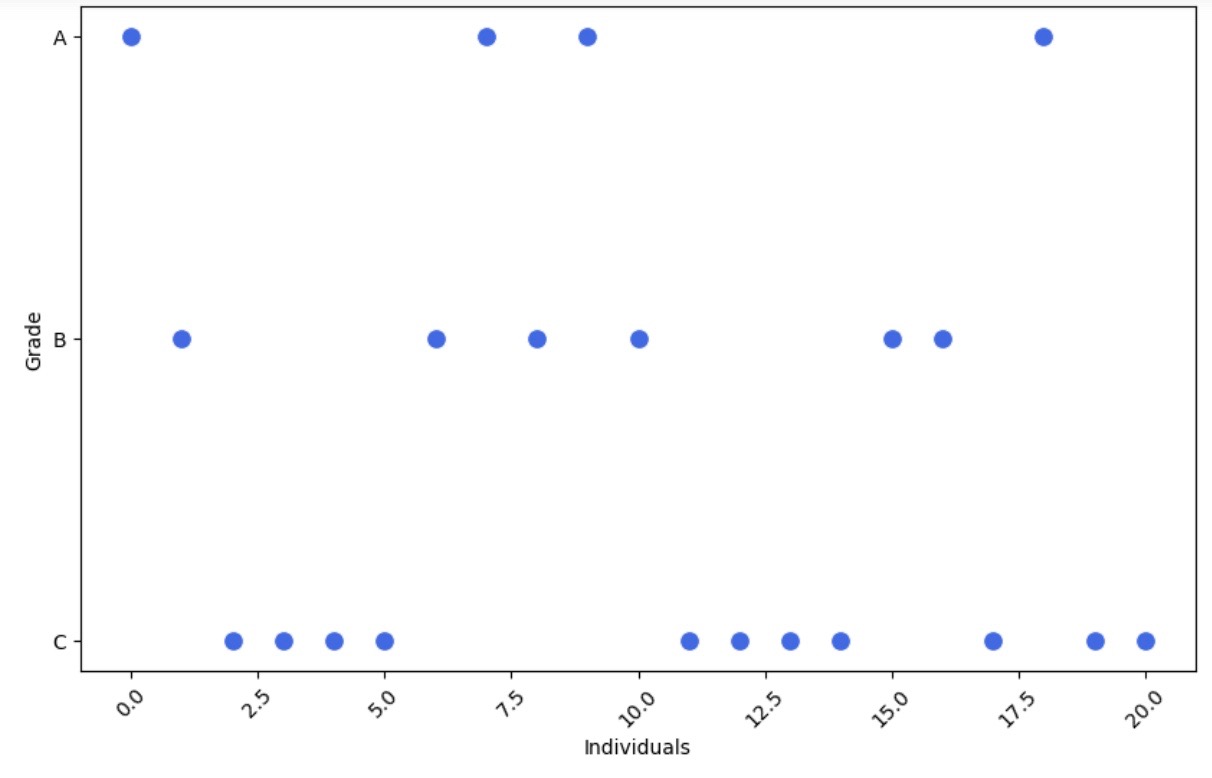}
        \label{fig:sub9}
    \end{minipage}
    \hfill
    \begin{minipage}[b]{0.45\textwidth}
        \centering
        \includegraphics[width=\textwidth]{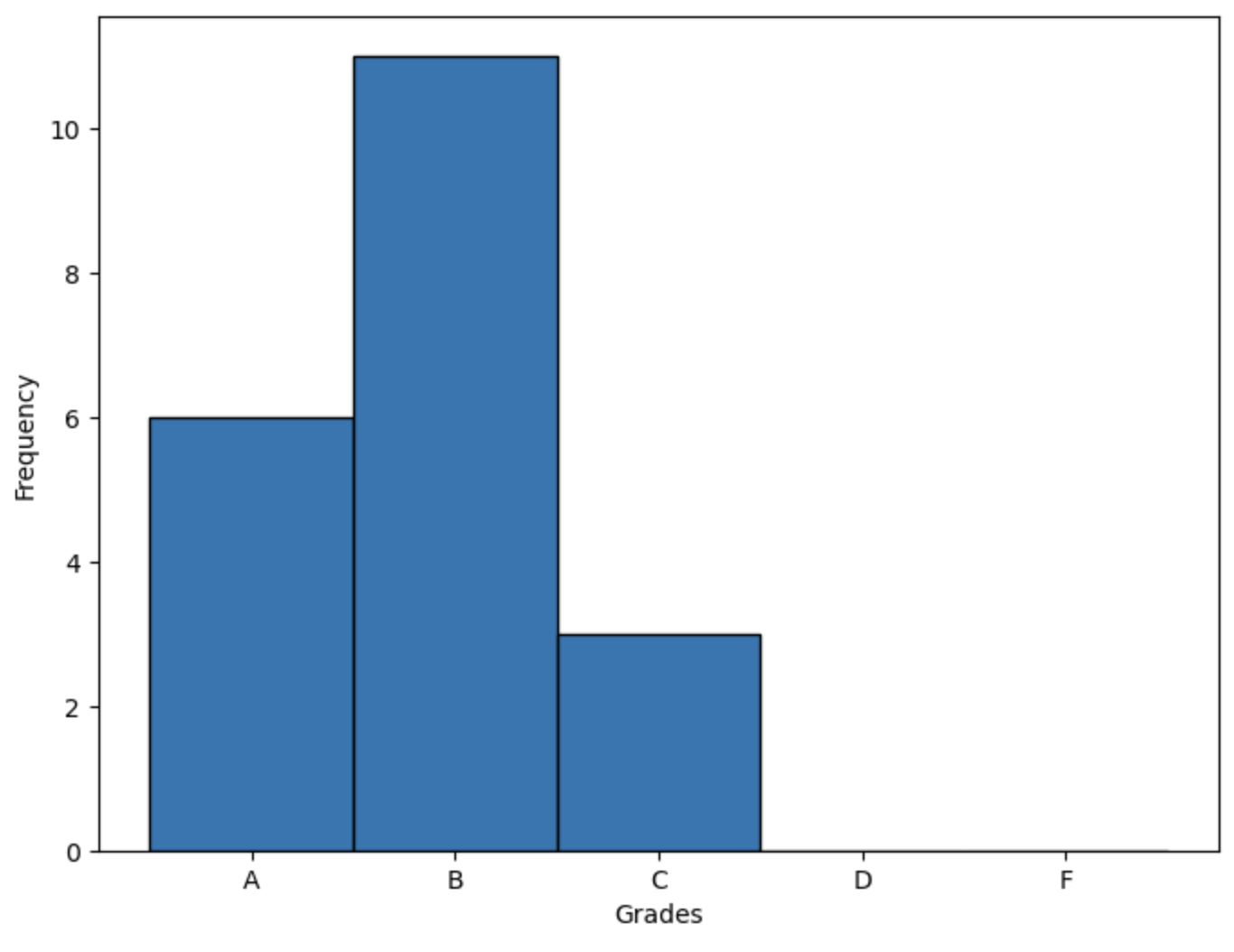}
        \label{fig:sub9b}
    \end{minipage}
    \hfill
    \begin{minipage}[b]{0.45\textwidth}
        \centering
        \includegraphics[width=\textwidth]{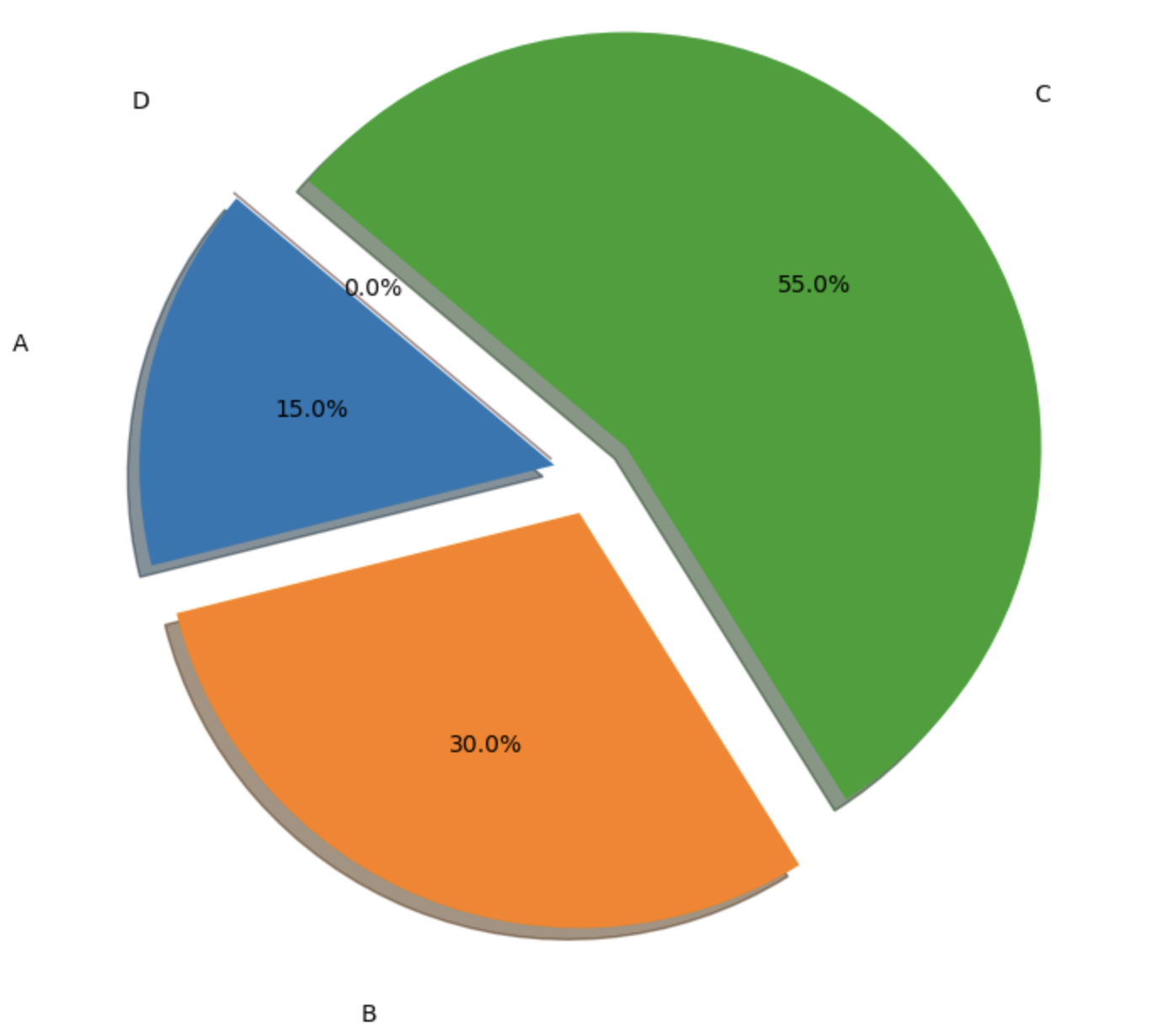}
        \label{fig:sub9c}
    \end{minipage}
    \caption{ISTJ grade distribution graph, a) The scatter of ISTJ grade distribution in Department of Physics, Zhejiang University, b) The histogram of ISTJ personality grade distribution in Department of Physics, Zhejiang University c)The pie chart of ISTJ personality grade distribution in Department of Physics, Zhejiang University}
    \label{fig:istj}
\end{figure}

\subsection{Remarkable MBTI personality analysis : ISTJ}
Last but not least, ISTJs are important in academic cycle. The analysis of academic performance among ISTJ students reveals distinct patterns in their grading distribution.   The A grade category includes 4 individuals (approximately 15.0\%), indicating a modest level of high achievers within the ISTJ population (Shown in Figure \ref{fig:istj}).  This suggests that while some ISTJs excel academically, they do not represent the majority in this group. The B grade category shows 6 individuals (about 30.0\%), reflecting a similar level of academic performance as those achieving A grades, indicating that a balanced number of ISTJs are performing well.   However, the C grade category has the highest representation, with 11 individuals (approximately 55.0\%), suggesting that a notable portion of ISTJ students may face challenges in their academic performance.   This trend could reflect the ISTJ preference for structure and detail, which might only sometimes align with the requirements of more flexible or abstract learning environments \citep{maccann2020emotional}.

ISTJs are characterized by their meticulousness, practicality, and reliability, which are advantageous in academic settings where attention to detail and adherence to established protocols are crucial \citep{melear1989cognitive}. Their strong organizational skills enable them to develop structured study plans and effectively manage their time, essential qualities for navigating the demands of rigorous academic life. Additionally, ISTJs' commitment to their responsibilities often translates into a disciplined approach to learning, which can facilitate success in environments that value consistency and hard work\citep{stein2019evaluating}. However, the relatively equal distribution of A and B grades indicates that while there are some strong performers, a significant number of ISTJ students achieve moderate success.   This observation may highlight the urgent need for additional academic support for those struggling to maintain higher grades.   The presence of C grades emphasizes the potential difficulties that ISTJs might encounter, possibly due to their adherence to structured approaches that may not suit all subjects or teaching styles.   For instance, ISTJs may thrive in courses that emphasize clear guidelines and logical reasoning, yet they may struggle in more abstract or open-ended assignments that require creative thinking and flexibility \citep{markon2009hierarchies}.

While ISTJ students demonstrate solid academic performance overall, their distinct characteristics—such as a preference for structure, reliability, and strong organizational skills—position them well for success in academic environments that value these traits. Nevertheless, educational institutions should consider implementing support systems that help ISTJ students navigate subjects that challenge their structured learning style, ultimately enhancing their academic experience and potential for growth and achievement in academia. By recognizing and accommodating ISTJs' unique strengths and challenges, educators can foster an environment conducive to their growth and achievement.

\begin{figure}[h]
    \centering
    \includegraphics[width=0.8\linewidth]{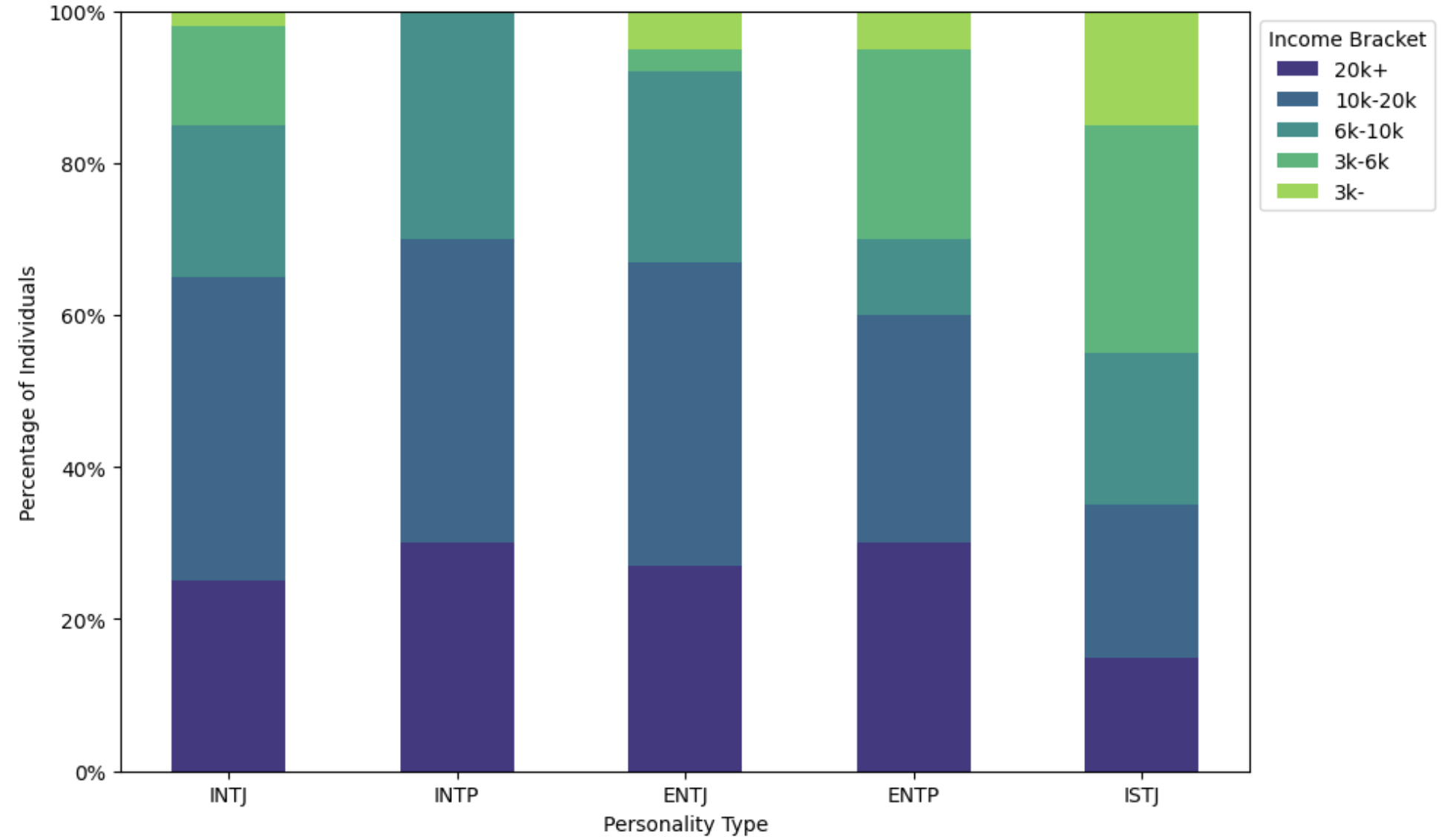}
    \caption{Family monthly income per person with respect to each MBTI personality}
    \label{fig:income}
\end{figure}
\subsection{Remarkable MBTI personality analysis : Income distribution}
When discussing the correlation between MBTI and an individual's household income. The analysis of parental income distribution across various Myers-Briggs Type Indicator (MBTI) personality types has unveiled significant correlations. These correlations provide intriguing insights into the socioeconomic backgrounds of individuals within each category, shedding light on how personality traits may correlate with financial outcomes and influence academic trajectories.

The INTJ personality type demonstrates a noteworthy association with higher parental income levels, as 25\% of individuals have parents earning ¥20,000 or more, indicating a significant presence of high earners among INTJs (Shown in Figure \ref{fig:income}). Additionally, 40\% of INTJs fall within the ¥10,000 to ¥20,000 parental income range, reflecting a stable financial standing. This financial stability results in supplying the structured, achievement-oriented environments in which many INTJs are raised, fostering their innate tendencies toward strategic thinking and long-term planning \citep{soliday1992study}. Nurturing these characteristics in supportive financial settings can cultivate academic success, allowing INTJs to leverage their analytical strengths in rigorous academic pursuits.

In contrast, the INTP personality type displays an even higher percentage of individuals with parents earning in the ¥20,000 or more bracket, with 30\% achieving this income level (Shown in Figure \ref{fig:income}). Similar to INTJs, 40\% of INTPs have parents in the ¥10,000 to 20,000 range, suggesting a robust financial profile. The prevalence of these parental income levels among INTPs indicates that their family environments likely encourage exploration and intellectual curiosity, aligning with their strengths in problem-solving and abstract thinking \citep{jach2023personality}. This nurturing environment can significantly enhance their academic trajectories, as INTPs thrive in settings that allow for creative exploration and critical inquiry.

For the ENTJ personality type, 27\% of individuals come from families earning ¥20,000 or more, while 40\% have parents in the ¥10,000 to ¥20,000 range (Shown in Figure \ref{fig:income}). The high concentration in upper-income brackets reflects the ambitious and driven nature of ENTJs, often nurtured in environments that emphasize leadership and achievement. Their upbringing likely encourages goal-oriented behaviours, which can translate into academic and professional success \citep{markon2009hierarchies}. The influence of family dynamics and financial resources thus plays a critical role in shaping ENTJ traits, enabling them to excel in academic and leadership roles.

The ENTP personality type presents a balanced distribution, with 30\% having parents earning ¥20,000 or more and another 30\% within the ¥10,000 to ¥20,000 bracket (Shown in Figure \ref{fig:income}). The presence of 25\% in the ¥3,000 to ¥6,000 range and 15\% in the lowest category indicates some economic diversity. ENTPs often thrive in environments that promote innovation and adaptability, suggesting that supportive family settings are crucial for cultivating their strengths. Families encouraging open-mindedness and creative thinking can better equip ENTPs for academic challenges requiring dynamic problem-solving skills \citep{barrett1985personality}.

Lastly, the ISTJ personality type shows that 15\% come from families earning ¥20,000 or more, with 20\% in the ¥10,000 to ¥20,000 range (Shown in Figure \ref{fig:income}). However, the most significant portion of ISTJs, 30\%, has parents earning between ¥3,000 to ¥6,000, reflecting a more conservative income distribution compared to other personality types. The influence of a structured and reliable family environment is evident here, as ISTJs are often raised in settings that prioritize order and discipline. This upbringing can help them develop strong academic skills, despite those from lower-income backgrounds may face obstacles in accessing resources necessary for academic advancement \citep{markon2009hierarchies}.

The data reveals distinct patterns in parental income distribution among different MBTI personality types, underscoring the significant influence of family environments. These environments play a crucial role in the development of personality traits that predispose individuals to succeed in academia. INTJs and INTPs, for example, are more likely to come from stable financial backgrounds that cultivate their intellectual capabilities, while ENTJs and ENTPs benefit from environments that foster leadership and innovation.

As a matter of fact, educational institutions should focus on creating supportive environments that nurture diverse personality traits to cultivate more individuals suited for academic pursuits. Strategies could include mentorship programs that connect students with role models in academia, workshops that foster critical thinking and creativity, and resources aimed at enhancing accessibility for students from lower-income backgrounds. By recognizing and accommodating the unique strengths of each personality type, educators can better prepare students for successful academic and professional careers, ultimately contributing to a more diverse and dynamic academic community.

\begin{figure}[ht]
    \centering
    \begin{minipage}[b]{0.6\textwidth}
        \centering
        \includegraphics[width=\textwidth]{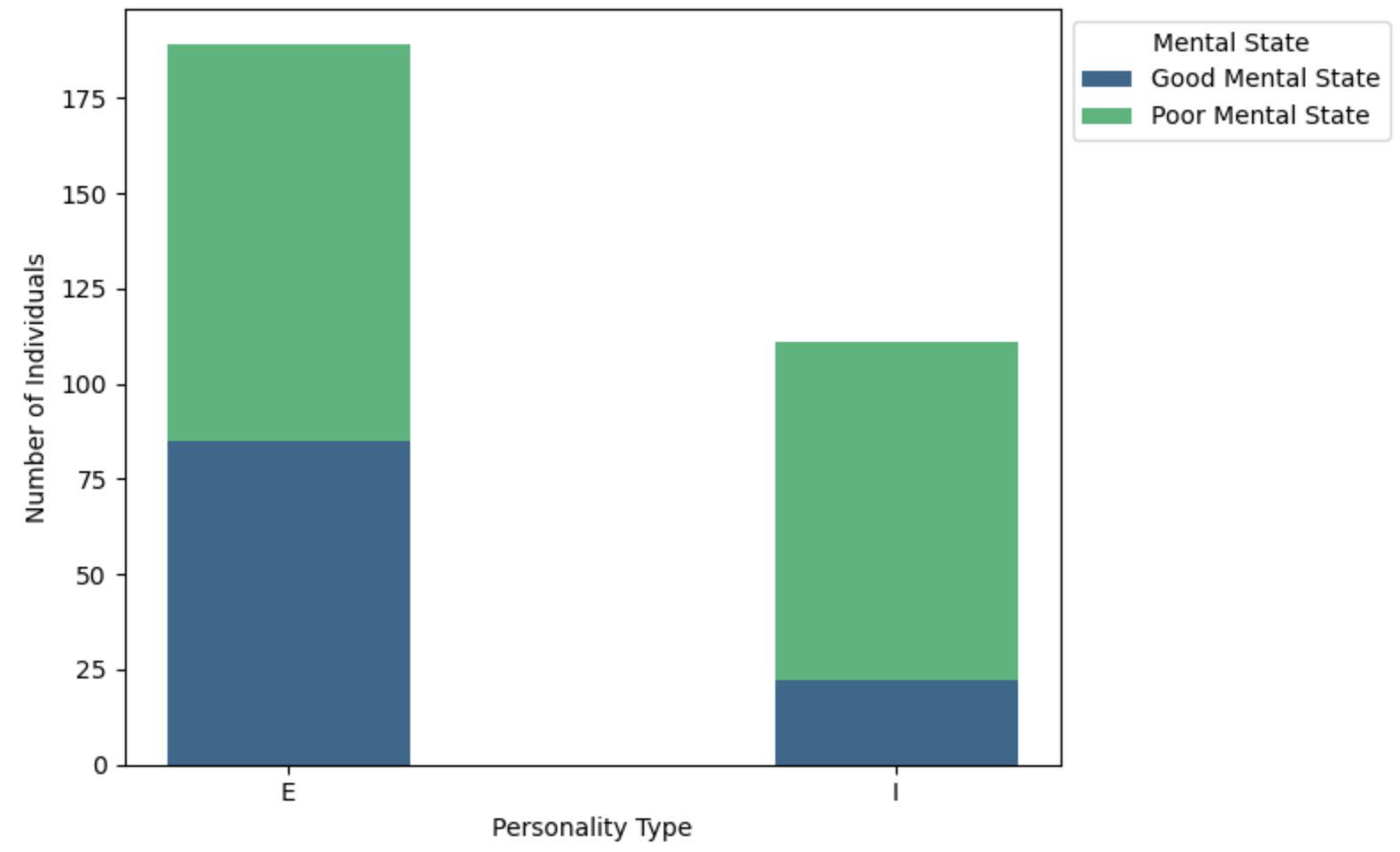}
        \label{fig:sub12}
    \end{minipage}
    \hfill
    \begin{minipage}[b]{0.6\textwidth}
        \centering
        \includegraphics[width=\textwidth]{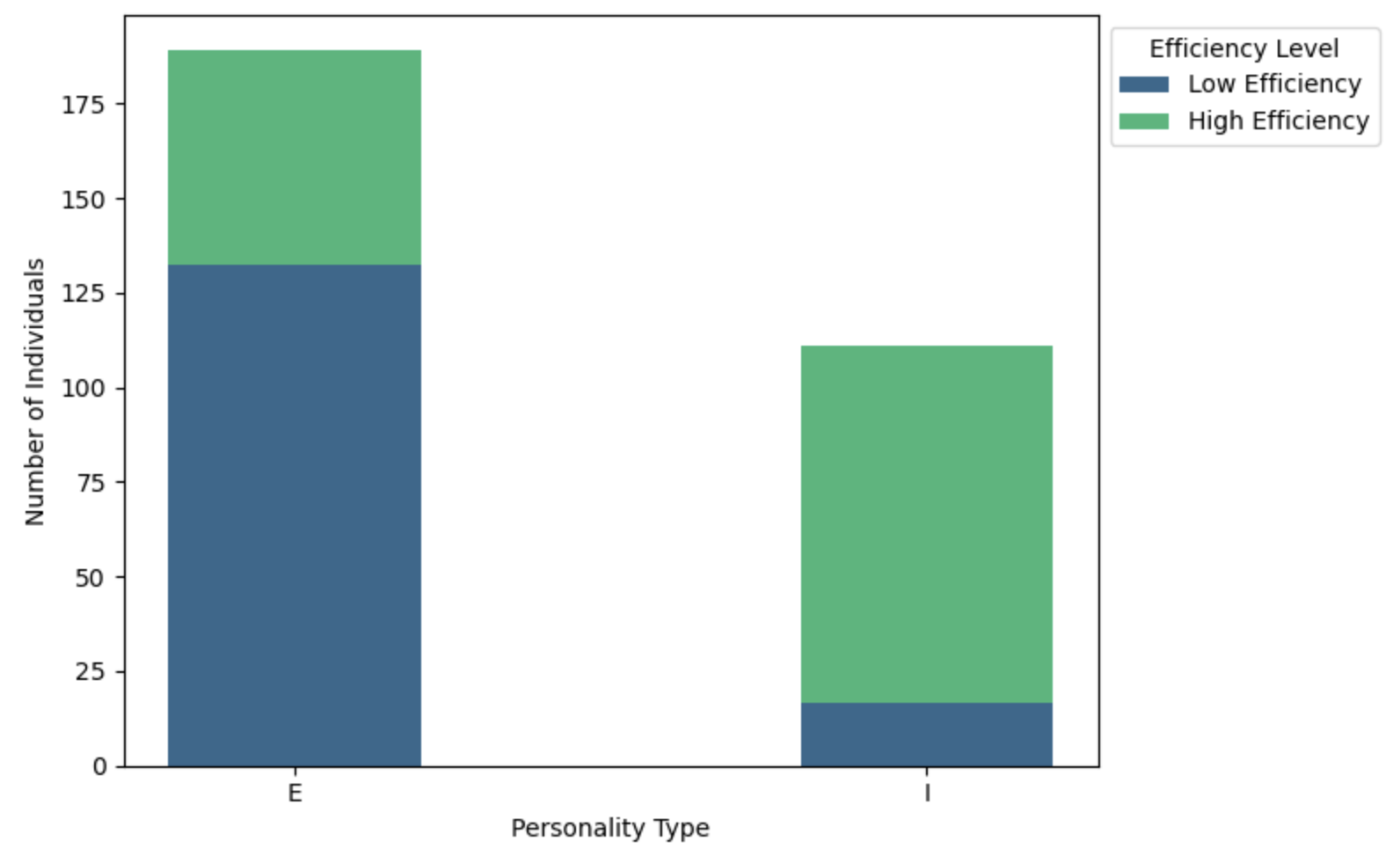}
        \label{fig:sub12b}
    \end{minipage}
    \caption{The histogram of extrovert-introvert(E-I) mental health and working efficiency among 300 samples: a) the histogram of extrovert-introvert(E-I) mental health among 300 samples, x axis is the personality type, y axis is the population, b) the histogram of extrovert-introvert(E-I) working efficiency among 300 samples, x axis is the personality type, y axis is the population. }
    \label{fig:12}
\end{figure}
Another important apesct that need to be mentioned is the interplay between personality traits, particularly introversion and extraversion, and mental stress is a critical area of study, especially concerning academic environments. Our research indicates that introverted individuals experience internal conflict more than extroverted individuals (shown in Figure \ref{fig:sub12}). However, extroverts may struggle with mental exhaustion, which can negatively impact their psychological health and work efficiency. This distinction suggests that introverted individuals often exhibit greater resilience under pressure, maintaining productivity even in the face of stress \citep{furnham2013personality}.

The findings also highlight that extroverted individuals typically seek various outlets to alleviate their stress, aiming to restore their optimal work state, albeit at the potential cost of time efficiency. In contrast, introverted individuals may suppress their mental strain, which can culminate in a sudden psychological breakdown that impairs their ability to work. Such dynamics underscore the importance of understanding how personality traits influence stress management strategies, particularly in academic settings where the pressures to perform are heightened.
Furthermore, our research underscores the significant role of financial stability in mitigating mental stress for both introverted and extroverted individuals. Parental income levels have a profound impact on the resources available to support students' mental health, shaping their coping mechanisms in high-pressure academic environments \citep{bradley2002socioeconomic}. Higher household income can provide access to mental health resources, extracurricular activities, and supportive educational environments, which are crucial for fostering resilience among students.

Educational institutions must consider several strategies to cultivate more individuals suited for academic pursuits and create environments conducive to their success. First, implementing mentorship programs can help students develop effective coping strategies tailored to their personality types. By pairing students with mentors who understand the unique challenges introverts and extroverts face, institutions can foster resilience and adaptability \citep{murphy2020relationship}. Second, universities should promote mental health awareness and provide accessible resources, such as counselling services, stress management workshops, and peer support groups. These initiatives can help students develop healthier coping mechanisms, reducing the likelihood of burnout and promoting sustained academic engagement.

Finally, it is crucial for educational institutions to foster an inclusive and supportive academic culture that values diverse personality traits. By creating environments where both introverts and extroverts feel valued and understood, institutions can encourage students to thrive. This approach not only enhances students' overall well-being and academic performance but also promotes a positive learning environment for all. It is essential for universities to understand the relationship among parental income, personality traits, and mental stress. This understanding can serve as a basis for developing effective strategies to support academic success. By addressing the unique needs of both introverted and extroverted students, institutions can cultivate a more resilient and capable academic community, ultimately contributing to the advancement of knowledge and innovation in various fields.

\section{Conclusion}\label{Conclusion}

This study comprehensively examines the distribution of Myers-Briggs Type Indicator (MBTI) personality types among physics undergraduates at Zhejiang University, alongside an analysis of their career aspirations, family income, and mental health status. The results indicate a significant correlation between personality traits and academic performance, revealing that INTJ and INTP students exhibit the strongest tendencies toward pursuing academic research, typically emerging from financially stable, middle- to upper-class backgrounds. This finding suggests that socioeconomic factors critically shape the academic trajectories of students, emphasizing the role of a supportive family environment in fostering intellectual curiosity and resilience.
The data indicate that introverted individuals, particularly those with INTJ and INTP personality types, tend to manage stress more effectively and maintain higher productivity levels in academic settings. Their analytical strengths and a preference for strategic planning and problem-solving enable them to navigate complex subjects with relative ease. In contrast, extroverted individuals are more susceptible to mental exhaustion, which can adversely affect their psychological health and work efficiency. The need for external outlets to alleviate stress often leads extroverted students to expend time and energy on stress-relief activities, which, while beneficial for their mental health, can detract from their academic focus and efficiency. This dynamic suggests that introverted students exhibit greater resilience under pressure. However, the potential for burnout among extroverted individuals raises concerns about their overall academic engagement and long-term success.

Furthermore, the analysis of parental income levels reveals that financial stability significantly mitigates mental stress for introverted and extroverted students. Higher household income correlates with improved access to mental health resources, extracurricular activities, and supportive educational environments. These resources are essential for fostering resilience and promoting sustained academic success. The results underscore the need for educational institutions to recognize the impact of socioeconomic backgrounds on student experiences, particularly in terms of accessing opportunities that enhance learning and well-being.

Educational institutions should implement targeted strategies that address the unique needs of their diverse student populations to cultivate a more robust academic community. First, mentorship programs that connect students with faculty and professionals who understand the specific challenges introverted and extroverted individuals face can provide invaluable guidance and support. Such relationships can foster resilience, promote effective coping strategies, and enhance students' academic engagement. Additionally, universities should prioritize mental health awareness by offering accessible resources such as counseling services, stress management workshops, and peer support groups. These initiatives can help students develop healthier coping mechanisms, reducing the likelihood of burnout and enhancing their ability to thrive in high-pressure academic environments. Creating a culture that normalizes discussions about mental health and emphasizes the importance of well-being will encourage students to seek help when needed.

Moreover, fostering an inclusive and supportive academic culture that values diverse personality traits is crucial. Institutions should create environments where introverts and extroverts feel valued and understood. This can be achieved through curricular and extracurricular activities promoting collaboration, creativity, and open-mindedness, allowing students to leverage their unique strengths in academic and social settings. By recognizing and accommodating the diverse needs of their students, educational institutions can enhance overall well-being and academic performance.

In conclusion, this study highlights the intricate relationship between personality traits, parental income, and mental stress in shaping the academic success of students. Addressing the unique needs of both introverted and extroverted individuals is essential for fostering a thriving academic community. By implementing tailored support systems and promoting an inclusive educational environment, institutions can empower students to excel academically and contribute meaningfully to their fields. Future research should continue to explore these dynamics, providing deeper insights into how we can better support the diverse needs of students in academia and enhance their potential for success in both academic and professional contexts.

\section*{Acknowledgments}
The author(s) declare that there is no conflict of interest regarding the publication of this paper.

\bibliographystyle{apsrev4-1}

\bibliography{oja_template}

\begin{thebibliography}{33}%
\makeatletter
\providecommand \@ifxundefined [1]{%
 \@ifx{#1\undefined}
}%
\providecommand \@ifnum [1]{%
 \ifnum #1\expandafter \@firstoftwo
 \else \expandafter \@secondoftwo
 \fi
}%
\providecommand \@ifx [1]{%
 \ifx #1\expandafter \@firstoftwo
 \else \expandafter \@secondoftwo
 \fi
}%
\providecommand \natexlab [1]{#1}%
\providecommand \enquote  [1]{``#1''}%
\providecommand \bibnamefont  [1]{#1}%
\providecommand \bibfnamefont [1]{#1}%
\providecommand \citenamefont [1]{#1}%
\providecommand \href@noop [0]{\@secondoftwo}%
\providecommand \href [0]{\begingroup \@sanitize@url \@href}%
\providecommand \@href[1]{\@@startlink{#1}\@@href}%
\providecommand \@@href[1]{\endgroup#1\@@endlink}%
\providecommand \@sanitize@url [0]{\catcode `\\12\catcode `\$12\catcode `\&12\catcode `\#12\catcode `\^12\catcode `\_12\catcode `\%12\relax}%
\providecommand \@@startlink[1]{}%
\providecommand \@@endlink[0]{}%
\providecommand \url  [0]{\begingroup\@sanitize@url \@url }%
\providecommand \@url [1]{\endgroup\@href {#1}{\urlprefix }}%
\providecommand \urlprefix  [0]{URL }%
\providecommand \Eprint [0]{\href }%
\providecommand \doibase [0]{http://dx.doi.org/}%
\providecommand \selectlanguage [0]{\@gobble}%
\providecommand \bibinfo  [0]{\@secondoftwo}%
\providecommand \bibfield  [0]{\@secondoftwo}%
\providecommand \translation [1]{[#1]}%
\providecommand \BibitemOpen [0]{}%
\providecommand \bibitemStop [0]{}%
\providecommand \bibitemNoStop [0]{.\EOS\space}%
\providecommand \EOS [0]{\spacefactor3000\relax}%
\providecommand \BibitemShut  [1]{\csname bibitem#1\endcsname}%
\let\auto@bib@innerbib\@empty
\bibitem [{\citenamefont {Myers}\ \emph {et~al.}(1998)\citenamefont {Myers}, \citenamefont {McCaulley}, \citenamefont {Quenk},\ and\ \citenamefont {Hammer}}]{myers1998}%
  \BibitemOpen
  \bibfield  {author} {\bibinfo {author} {\bibfnamefont {I.~B.}\ \bibnamefont {Myers}}, \bibinfo {author} {\bibfnamefont {M.~H.}\ \bibnamefont {McCaulley}}, \bibinfo {author} {\bibfnamefont {N.~L.}\ \bibnamefont {Quenk}}, \ and\ \bibinfo {author} {\bibfnamefont {A.~L.}\ \bibnamefont {Hammer}},\ }\href@noop {} {\emph {\bibinfo {title} {MBTI Manual: A Guide to the Development and Use of the Myers-Briggs Type Indicator}}}\ (\bibinfo  {publisher} {Consulting Psychologists Press},\ \bibinfo {address} {Palo Alto, CA},\ \bibinfo {year} {1998})\BibitemShut {NoStop}%
\bibitem [{\citenamefont {Buss}(2019)}]{buss2019evolutionary}%
  \BibitemOpen
  \bibfield  {author} {\bibinfo {author} {\bibfnamefont {D.}~\bibnamefont {Buss}},\ }\href@noop {} {\emph {\bibinfo {title} {Evolutionary psychology: The new science of the mind}}}\ (\bibinfo  {publisher} {Routledge},\ \bibinfo {year} {2019})\BibitemShut {NoStop}%
\bibitem [{\citenamefont {McCrae}\ and\ \citenamefont {Costa}(2003)}]{mccrae2003}%
  \BibitemOpen
  \bibfield  {author} {\bibinfo {author} {\bibfnamefont {R.~R.}\ \bibnamefont {McCrae}}\ and\ \bibinfo {author} {\bibfnamefont {P.~T.}\ \bibnamefont {Costa}},\ }\href@noop {} {\emph {\bibinfo {title} {Personality in adulthood: A five-factor theory perspective}}}\ (\bibinfo  {publisher} {Guilford Press},\ \bibinfo {year} {2003})\BibitemShut {NoStop}%
\bibitem [{\citenamefont {Sirin}(2005)}]{sirin2005socioeconomic}%
  \BibitemOpen
  \bibfield  {author} {\bibinfo {author} {\bibfnamefont {S.~R.}\ \bibnamefont {Sirin}},\ }\href@noop {} {\bibfield  {journal} {\bibinfo  {journal} {Review of educational research}\ }\textbf {\bibinfo {volume} {75}},\ \bibinfo {pages} {417} (\bibinfo {year} {2005})}\BibitemShut {NoStop}%
\bibitem [{\citenamefont {McCormack}\ \emph {et~al.}(1990)\citenamefont {McCormack}, \citenamefont {Patterson}, \citenamefont {Ohlde}, \citenamefont {Garfield},\ and\ \citenamefont {Schauer}}]{mccormack1990mmpi}%
  \BibitemOpen
  \bibfield  {author} {\bibinfo {author} {\bibfnamefont {J.~K.}\ \bibnamefont {McCormack}}, \bibinfo {author} {\bibfnamefont {T.~W.}\ \bibnamefont {Patterson}}, \bibinfo {author} {\bibfnamefont {C.~D.}\ \bibnamefont {Ohlde}}, \bibinfo {author} {\bibfnamefont {N.~J.}\ \bibnamefont {Garfield}}, \ and\ \bibinfo {author} {\bibfnamefont {A.}~\bibnamefont {Schauer}},\ }\href {\doibase 10.1207/s15327752jpa5403\&4\_17} {\bibfield  {journal} {\bibinfo  {journal} {Journal of Personality Assessment}\ }\textbf {\bibinfo {volume} {54}},\ \bibinfo {pages} {628} (\bibinfo {year} {1990})}\BibitemShut {NoStop}%
\bibitem [{\citenamefont {Jung}(1971)}]{jung1971personality}%
  \BibitemOpen
  \bibfield  {author} {\bibinfo {author} {\bibfnamefont {C.~G.}\ \bibnamefont {Jung}},\ }\href@noop {} {\bibfield  {journal} {\bibinfo  {journal} {The portable Jung}\ ,\ \bibinfo {pages} {178}} (\bibinfo {year} {1971})}\BibitemShut {NoStop}%
\bibitem [{\citenamefont {Poropat}(2009)}]{poropat2009meta}%
  \BibitemOpen
  \bibfield  {author} {\bibinfo {author} {\bibfnamefont {A.~E.}\ \bibnamefont {Poropat}},\ }\href@noop {} {\bibfield  {journal} {\bibinfo  {journal} {Psychological bulletin}\ }\textbf {\bibinfo {volume} {135}},\ \bibinfo {pages} {322} (\bibinfo {year} {2009})}\BibitemShut {NoStop}%
\bibitem [{\citenamefont {Vesely}\ \emph {et~al.}(2014)\citenamefont {Vesely}, \citenamefont {Saklofske},\ and\ \citenamefont {Nordstokke}}]{vesely2014ei}%
  \BibitemOpen
  \bibfield  {author} {\bibinfo {author} {\bibfnamefont {A.~K.}\ \bibnamefont {Vesely}}, \bibinfo {author} {\bibfnamefont {D.~H.}\ \bibnamefont {Saklofske}}, \ and\ \bibinfo {author} {\bibfnamefont {D.~W.}\ \bibnamefont {Nordstokke}},\ }\href@noop {} {\bibfield  {journal} {\bibinfo  {journal} {Personality and individual differences}\ }\textbf {\bibinfo {volume} {65}},\ \bibinfo {pages} {81} (\bibinfo {year} {2014})}\BibitemShut {NoStop}%
\bibitem [{\citenamefont {Zhang}(2002)}]{zhang2002thinking}%
  \BibitemOpen
  \bibfield  {author} {\bibinfo {author} {\bibfnamefont {L.-F.}\ \bibnamefont {Zhang}},\ }\href@noop {} {\bibfield  {journal} {\bibinfo  {journal} {Educational psychology}\ }\textbf {\bibinfo {volume} {22}},\ \bibinfo {pages} {17} (\bibinfo {year} {2002})}\BibitemShut {NoStop}%
\bibitem [{\citenamefont {White}(1982)}]{white1982relation}%
  \BibitemOpen
  \bibfield  {author} {\bibinfo {author} {\bibfnamefont {K.~R.}\ \bibnamefont {White}},\ }\href@noop {} {\bibfield  {journal} {\bibinfo  {journal} {Psychological bulletin}\ }\textbf {\bibinfo {volume} {91}},\ \bibinfo {pages} {461} (\bibinfo {year} {1982})}\BibitemShut {NoStop}%
\bibitem [{\citenamefont {McLellan}\ and\ \citenamefont {Jackson}(2017)}]{mclellan2017personality}%
  \BibitemOpen
  \bibfield  {author} {\bibinfo {author} {\bibfnamefont {C.~K.}\ \bibnamefont {McLellan}}\ and\ \bibinfo {author} {\bibfnamefont {D.~L.}\ \bibnamefont {Jackson}},\ }\href@noop {} {\bibfield  {journal} {\bibinfo  {journal} {Social Psychology of Education}\ }\textbf {\bibinfo {volume} {20}},\ \bibinfo {pages} {159} (\bibinfo {year} {2017})}\BibitemShut {NoStop}%
\bibitem [{\citenamefont {Widiger}\ and\ \citenamefont {Costa~Jr}(2013)}]{widiger2013personality}%
  \BibitemOpen
  \bibfield  {author} {\bibinfo {author} {\bibfnamefont {T.~A.}\ \bibnamefont {Widiger}}\ and\ \bibinfo {author} {\bibfnamefont {P.~T.}\ \bibnamefont {Costa~Jr}},\ }\href@noop {} {\bibfield  {journal} {\bibinfo  {journal} {American Psychological Association: Washington, DC, USA}\ ,\ \bibinfo {pages} {3}} (\bibinfo {year} {2013})}\BibitemShut {NoStop}%
\bibitem [{\citenamefont {O’Connor}\ and\ \citenamefont {Paunonen}(2007)}]{o2007big}%
  \BibitemOpen
  \bibfield  {author} {\bibinfo {author} {\bibfnamefont {M.~C.}\ \bibnamefont {O’Connor}}\ and\ \bibinfo {author} {\bibfnamefont {S.~V.}\ \bibnamefont {Paunonen}},\ }\href@noop {} {\bibfield  {journal} {\bibinfo  {journal} {Personality and Individual differences}\ }\textbf {\bibinfo {volume} {43}},\ \bibinfo {pages} {971} (\bibinfo {year} {2007})}\BibitemShut {NoStop}%
\bibitem [{\citenamefont {Vesely}\ \emph {et~al.}(2013)\citenamefont {Vesely}, \citenamefont {Saklofske},\ and\ \citenamefont {Leschied}}]{vesely2013teachers}%
  \BibitemOpen
  \bibfield  {author} {\bibinfo {author} {\bibfnamefont {A.~K.}\ \bibnamefont {Vesely}}, \bibinfo {author} {\bibfnamefont {D.~H.}\ \bibnamefont {Saklofske}}, \ and\ \bibinfo {author} {\bibfnamefont {A.~D.}\ \bibnamefont {Leschied}},\ }\href@noop {} {\bibfield  {journal} {\bibinfo  {journal} {Canadian Journal of School Psychology}\ }\textbf {\bibinfo {volume} {28}},\ \bibinfo {pages} {71} (\bibinfo {year} {2013})}\BibitemShut {NoStop}%
\bibitem [{\citenamefont {Stein}\ and\ \citenamefont {Swan}(2019)}]{stein2019evaluating}%
  \BibitemOpen
  \bibfield  {author} {\bibinfo {author} {\bibfnamefont {R.}~\bibnamefont {Stein}}\ and\ \bibinfo {author} {\bibfnamefont {A.~B.}\ \bibnamefont {Swan}},\ }\href@noop {} {\bibfield  {journal} {\bibinfo  {journal} {Social and Personality Psychology Compass}\ }\textbf {\bibinfo {volume} {13}},\ \bibinfo {pages} {e12434} (\bibinfo {year} {2019})}\BibitemShut {NoStop}%
\bibitem [{\citenamefont {Komarraju}\ \emph {et~al.}(2011)\citenamefont {Komarraju}, \citenamefont {Karau}, \citenamefont {Schmeck},\ and\ \citenamefont {Avdic}}]{komarraju2011big}%
  \BibitemOpen
  \bibfield  {author} {\bibinfo {author} {\bibfnamefont {M.}~\bibnamefont {Komarraju}}, \bibinfo {author} {\bibfnamefont {S.~J.}\ \bibnamefont {Karau}}, \bibinfo {author} {\bibfnamefont {R.~R.}\ \bibnamefont {Schmeck}}, \ and\ \bibinfo {author} {\bibfnamefont {A.}~\bibnamefont {Avdic}},\ }\href@noop {} {\bibfield  {journal} {\bibinfo  {journal} {Personality and individual differences}\ }\textbf {\bibinfo {volume} {51}},\ \bibinfo {pages} {472} (\bibinfo {year} {2011})}\BibitemShut {NoStop}%
\bibitem [{\citenamefont {Richardson}\ \emph {et~al.}(2012)\citenamefont {Richardson}, \citenamefont {Abraham},\ and\ \citenamefont {Bond}}]{richardson2012psychological}%
  \BibitemOpen
  \bibfield  {author} {\bibinfo {author} {\bibfnamefont {M.}~\bibnamefont {Richardson}}, \bibinfo {author} {\bibfnamefont {C.}~\bibnamefont {Abraham}}, \ and\ \bibinfo {author} {\bibfnamefont {R.}~\bibnamefont {Bond}},\ }\href@noop {} {\bibfield  {journal} {\bibinfo  {journal} {Psychological bulletin}\ }\textbf {\bibinfo {volume} {138}},\ \bibinfo {pages} {353} (\bibinfo {year} {2012})}\BibitemShut {NoStop}%
\bibitem [{\citenamefont {Furnham}\ and\ \citenamefont {Chamorro-Premuzic}(2004)}]{furnham2004personality}%
  \BibitemOpen
  \bibfield  {author} {\bibinfo {author} {\bibfnamefont {A.}~\bibnamefont {Furnham}}\ and\ \bibinfo {author} {\bibfnamefont {T.}~\bibnamefont {Chamorro-Premuzic}},\ }\href@noop {} {\bibfield  {journal} {\bibinfo  {journal} {Personality and individual differences}\ }\textbf {\bibinfo {volume} {37}},\ \bibinfo {pages} {943} (\bibinfo {year} {2004})}\BibitemShut {NoStop}%
\bibitem [{\citenamefont {Matthews}\ \emph {et~al.}(2003)\citenamefont {Matthews}, \citenamefont {Deary},\ and\ \citenamefont {Whiteman}}]{matthews2003personality}%
  \BibitemOpen
  \bibfield  {author} {\bibinfo {author} {\bibfnamefont {G.}~\bibnamefont {Matthews}}, \bibinfo {author} {\bibfnamefont {I.~J.}\ \bibnamefont {Deary}}, \ and\ \bibinfo {author} {\bibfnamefont {M.~C.}\ \bibnamefont {Whiteman}},\ }\href@noop {} {\emph {\bibinfo {title} {Personality traits}}}\ (\bibinfo  {publisher} {Cambridge University Press},\ \bibinfo {year} {2003})\BibitemShut {NoStop}%
\bibitem [{\citenamefont {Smidt}(2015)}]{smidt2015big}%
  \BibitemOpen
  \bibfield  {author} {\bibinfo {author} {\bibfnamefont {W.}~\bibnamefont {Smidt}},\ }\href@noop {} {\bibfield  {journal} {\bibinfo  {journal} {Journal of Education for Teaching}\ }\textbf {\bibinfo {volume} {41}},\ \bibinfo {pages} {385} (\bibinfo {year} {2015})}\BibitemShut {NoStop}%
\bibitem [{\citenamefont {Heimlich}(1990)}]{heimlich1990measuring}%
  \BibitemOpen
  \bibfield  {author} {\bibinfo {author} {\bibfnamefont {J.~E.}\ \bibnamefont {Heimlich}},\ }\href@noop {} {\emph {\bibinfo {title} {Measuring teaching style: A correlational study between the Van Tilburg/Heimlich Sensitivity Measure and the Myers-Briggs personality indicator on adult educators in central Ohio}}}\ (\bibinfo  {publisher} {The Ohio State University},\ \bibinfo {year} {1990})\BibitemShut {NoStop}%
\bibitem [{\citenamefont {Chamorro-Premuzic}\ and\ \citenamefont {Furnham}(2003)}]{chamorro2003personality}%
  \BibitemOpen
  \bibfield  {author} {\bibinfo {author} {\bibfnamefont {T.}~\bibnamefont {Chamorro-Premuzic}}\ and\ \bibinfo {author} {\bibfnamefont {A.}~\bibnamefont {Furnham}},\ }\href@noop {} {\bibfield  {journal} {\bibinfo  {journal} {European journal of Personality}\ }\textbf {\bibinfo {volume} {17}},\ \bibinfo {pages} {237} (\bibinfo {year} {2003})}\BibitemShut {NoStop}%
\bibitem [{\citenamefont {Vacha-Haase}(1998)}]{vacha1998reliability}%
  \BibitemOpen
  \bibfield  {author} {\bibinfo {author} {\bibfnamefont {T.}~\bibnamefont {Vacha-Haase}},\ }\href@noop {} {\bibfield  {journal} {\bibinfo  {journal} {Educational and Psychological Measurement}\ }\textbf {\bibinfo {volume} {58}},\ \bibinfo {pages} {6} (\bibinfo {year} {1998})}\BibitemShut {NoStop}%
\bibitem [{\citenamefont {Butterfuss}\ and\ \citenamefont {Kendeou}(2018)}]{butterfuss2018role}%
  \BibitemOpen
  \bibfield  {author} {\bibinfo {author} {\bibfnamefont {R.}~\bibnamefont {Butterfuss}}\ and\ \bibinfo {author} {\bibfnamefont {P.}~\bibnamefont {Kendeou}},\ }\href@noop {} {\bibfield  {journal} {\bibinfo  {journal} {Educational Psychology Review}\ }\textbf {\bibinfo {volume} {30}},\ \bibinfo {pages} {801} (\bibinfo {year} {2018})}\BibitemShut {NoStop}%
\bibitem [{\citenamefont {MacCann}\ \emph {et~al.}(2020)\citenamefont {MacCann}, \citenamefont {Jiang}, \citenamefont {Brown}, \citenamefont {Double}, \citenamefont {Bucich},\ and\ \citenamefont {Minbashian}}]{maccann2020emotional}%
  \BibitemOpen
  \bibfield  {author} {\bibinfo {author} {\bibfnamefont {C.}~\bibnamefont {MacCann}}, \bibinfo {author} {\bibfnamefont {Y.}~\bibnamefont {Jiang}}, \bibinfo {author} {\bibfnamefont {L.~E.}\ \bibnamefont {Brown}}, \bibinfo {author} {\bibfnamefont {K.~S.}\ \bibnamefont {Double}}, \bibinfo {author} {\bibfnamefont {M.}~\bibnamefont {Bucich}}, \ and\ \bibinfo {author} {\bibfnamefont {A.}~\bibnamefont {Minbashian}},\ }\href@noop {} {\bibfield  {journal} {\bibinfo  {journal} {Psychological bulletin}\ }\textbf {\bibinfo {volume} {146}},\ \bibinfo {pages} {150} (\bibinfo {year} {2020})}\BibitemShut {NoStop}%
\bibitem [{\citenamefont {Melear}(1989)}]{melear1989cognitive}%
  \BibitemOpen
  \bibfield  {author} {\bibinfo {author} {\bibfnamefont {C.~T.}\ \bibnamefont {Melear}},\ }\href@noop {} {\emph {\bibinfo {title} {Cognitive processes in the Curry learning style framework as measured by the learning style profile and the Myers-Briggs Type Indicator among non-majors in college biology}}}\ (\bibinfo  {publisher} {The Ohio State University},\ \bibinfo {year} {1989})\BibitemShut {NoStop}%
\bibitem [{\citenamefont {Markon}(2009)}]{markon2009hierarchies}%
  \BibitemOpen
  \bibfield  {author} {\bibinfo {author} {\bibfnamefont {K.~E.}\ \bibnamefont {Markon}},\ }\href@noop {} {\bibfield  {journal} {\bibinfo  {journal} {Social and Personality Psychology Compass}\ }\textbf {\bibinfo {volume} {3}},\ \bibinfo {pages} {812} (\bibinfo {year} {2009})}\BibitemShut {NoStop}%
\bibitem [{\citenamefont {Soliday}(1992)}]{soliday1992study}%
  \BibitemOpen
  \bibfield  {author} {\bibinfo {author} {\bibfnamefont {S.~F.~L.}\ \bibnamefont {Soliday}},\ }\href@noop {} {\emph {\bibinfo {title} {A study of personality types/learning styles of secondary vocational-technical education students}}}\ (\bibinfo  {publisher} {Oklahoma State University},\ \bibinfo {year} {1992})\BibitemShut {NoStop}%
\bibitem [{\citenamefont {Jach}\ \emph {et~al.}(2023)\citenamefont {Jach}, \citenamefont {Bardach},\ and\ \citenamefont {Murayama}}]{jach2023personality}%
  \BibitemOpen
  \bibfield  {author} {\bibinfo {author} {\bibfnamefont {H.~K.}\ \bibnamefont {Jach}}, \bibinfo {author} {\bibfnamefont {L.}~\bibnamefont {Bardach}}, \ and\ \bibinfo {author} {\bibfnamefont {K.}~\bibnamefont {Murayama}},\ }\href@noop {} {\bibfield  {journal} {\bibinfo  {journal} {Educational Psychology Review}\ }\textbf {\bibinfo {volume} {35}},\ \bibinfo {pages} {94} (\bibinfo {year} {2023})}\BibitemShut {NoStop}%
\bibitem [{\citenamefont {Barrett}(1985)}]{barrett1985personality}%
  \BibitemOpen
  \bibfield  {author} {\bibinfo {author} {\bibfnamefont {L.}~\bibnamefont {Barrett}},\ }\href@noop {} {\bibfield  {journal} {\bibinfo  {journal} {Journal of Agricultural Education}\ }\textbf {\bibinfo {volume} {26}},\ \bibinfo {pages} {48} (\bibinfo {year} {1985})}\BibitemShut {NoStop}%
\bibitem [{\citenamefont {Furnham}\ \emph {et~al.}(2013)\citenamefont {Furnham}, \citenamefont {Nuygards},\ and\ \citenamefont {Chamorro-Premuzic}}]{furnham2013personality}%
  \BibitemOpen
  \bibfield  {author} {\bibinfo {author} {\bibfnamefont {A.}~\bibnamefont {Furnham}}, \bibinfo {author} {\bibfnamefont {S.}~\bibnamefont {Nuygards}}, \ and\ \bibinfo {author} {\bibfnamefont {T.}~\bibnamefont {Chamorro-Premuzic}},\ }\href@noop {} {\bibfield  {journal} {\bibinfo  {journal} {Instructional science}\ }\textbf {\bibinfo {volume} {41}},\ \bibinfo {pages} {975} (\bibinfo {year} {2013})}\BibitemShut {NoStop}%
\bibitem [{\citenamefont {Bradley}\ and\ \citenamefont {Corwyn}(2002)}]{bradley2002socioeconomic}%
  \BibitemOpen
  \bibfield  {author} {\bibinfo {author} {\bibfnamefont {R.~H.}\ \bibnamefont {Bradley}}\ and\ \bibinfo {author} {\bibfnamefont {R.~F.}\ \bibnamefont {Corwyn}},\ }\href@noop {} {\bibfield  {journal} {\bibinfo  {journal} {Annual review of psychology}\ }\textbf {\bibinfo {volume} {53}},\ \bibinfo {pages} {371} (\bibinfo {year} {2002})}\BibitemShut {NoStop}%
\bibitem [{\citenamefont {Murphy}\ \emph {et~al.}(2020)\citenamefont {Murphy}, \citenamefont {Eduljee}, \citenamefont {Croteau},\ and\ \citenamefont {Parkman}}]{murphy2020relationship}%
  \BibitemOpen
  \bibfield  {author} {\bibinfo {author} {\bibfnamefont {L.}~\bibnamefont {Murphy}}, \bibinfo {author} {\bibfnamefont {N.~B.}\ \bibnamefont {Eduljee}}, \bibinfo {author} {\bibfnamefont {K.}~\bibnamefont {Croteau}}, \ and\ \bibinfo {author} {\bibfnamefont {S.}~\bibnamefont {Parkman}},\ }\href@noop {} {\bibfield  {journal} {\bibinfo  {journal} {International Journal of Research in Education and Science}\ }\textbf {\bibinfo {volume} {6}},\ \bibinfo {pages} {100} (\bibinfo {year} {2020})}\BibitemShut {NoStop}%
\end{thebibliography}%

\begin{appendix}

\section{Appendix 1: Questionnaire}
\label{ap:ap}
\noindent Choose (Strongly Disagree) -10 to 10 (Strongly Agree), with 0 indicating neutrality. 

\begin{itemize}
    \item \textbf{Gender:} \_\_\_\_\_\_
    \item \textbf{Age:} \_\_\_\_\_\_
    \item \textbf{Score:} \_\_\_\_\_\_
\end{itemize}

\begin{enumerate}
    \item I am a goal-oriented person.
    \item I am good at making plans, but I often fail to execute them.
    \item I typically focus more on the benefits I can gain from doing something rather than the reasons for doing it.
    \item When others do things I cannot understand, I often try to figure out their reasons for doing so.
    \item I evaluate a person's success based on the current societal value judgment system.
    \item I am a highly effective executor; I can always successfully plan and execute various tasks.
    \item When I am learning new knowledge, I am always thinking about how I should apply it.
    \item I often feel that I have a mission to promote the development of humanity in a humanitarian direction.
    \item My emotions are so abundant that even small daily occurrences, such as rain or falling leaves, can stir me deeply.
    \item I have clear likes and dislikes, and no event or person can keep me completely neutral.
    \item I enjoy harmonious interpersonal relationships.
    \item I never feel ashamed to express my emotional needs in front of close individuals.
    \item When a close friend cries in front of me, my first reaction is to think about what I should do to comfort them, rather than sincerely empathizing with them or considering how I can help.
    \item I always strive to meet the expectations others have of me.
    \item I believe I am the most unique person in the world.
    \item I enjoy watching extreme sports shows or participating in extreme sports (e.g., racing, skiing, skydiving, wingsuit flying, motorcycling).
    \item I am a nostalgic person; I like to keep old items/photos and can always recall warm and beautiful memories from them.
    \item I enjoy and excel at playing scenario simulation competitive games (e.g., Call of Duty, CS: GO, GTA5, Peace Elite).
    \item I am detail-oriented and can always notice very small details happening around me, such as a classmate's glasses nose pads falling off.
    \item When I cross the street and a car approaches, I can react faster than others.
    \item I trust authority and believe that people should act according to certain social rules.
    \item I hope to live in an orderly and methodical way.
    \item My thoughts are profound, and I can often see the essence of things before others.
    \item I have diverse interests, and my friends say I am well-educated.
    \item When learning new knowledge, I prefer to understand the knowledge itself systematically rather than its position or role within the overall knowledge system.
    \item I always enjoy reading obscure and difficult books.
    \item I frequently generate novel ideas and record them for future reference.
    \item I am a person who gets easily bored; when something loses its novelty, it becomes insignificant to me.
    \item My thinking is as vast as the ocean, filled with an uncontrollable curiosity about the world and a desire to explore the unknown.
    \item I am a person with high moral standards; no matter what happens, I will not cross my moral bottom line to harm others or use them for my benefit.
    \item For the ideals I hold, I do not consider the moral bottom line to be of great importance.
    \item To achieve my goals, I am willing to do things I greatly dislike.
    \item If I feel out of place in this world, I hope to stay true to myself until death.
    \item My life motto is to enjoy the moment; I never think about what will happen in the future.
    \item I always use my past experiences to plan for the futures of those around me.
    \item I often do many things that I believe are beneficial to others.
    \item I always pay attention to the willingness of those around me to do what I am about to do.
    \item I care about how others evaluate me.
    \item I believe love should be passionate and intense, and that my love must be romantic and full of passion.
    \item I always like to teach others how to do things correctly based on my past successful experiences.
    \item I have good taste and know how to maximize my enjoyment of life (e.g., I know the best places to eat and have fun in my area and what brands of lotion or shampoo work best).
    \item I am a rule-following person and cannot do things that violate social rules.
    \item As I age and gain experience, I increasingly find my previous thoughts to be naive and foolish.
    \item I have deep feelings for this world and often critique its problems.
    \item I focus on my life and rarely pay attention to matters unrelated to myself.
    \item I believe the abilities of my family and friends are more trustworthy than those of others.
    \item Although I strongly dislike others questioning my decisions, I can still listen to and accept their reasonable suggestions and opinions rationally.
    \item Pleasant weather always puts me in a good mood.
    \item If I do not need to go out, I can spend 2-3 days indoors without feeling suffocated.
    \item When a task needs to be completed, I tend to start working on it only as the deadline approaches.
    \item I detest chaos, as it irritates me and affects my efficiency.
    \item I occasionally do things that others consider unconventional.
    \item I believe that financial insufficiency has a significant impact on my motivation to study.
    \item What are your plans after graduation? Do you intend to pursue a PhD? Are you willing to become a scientist?
    \item What is your family's annual income in thousands of RMB?
    \item How many hours do you sleep each night?
    \item Do you often feel anxious?
    \item How many hours do you study each week?
    \item Do you participate in social activities, sports, or competitive gaming events?
    \item How many close friends do you have?
    \item Are you a person with diverse interests?
    \item Do you often feel physical discomfort, such as headaches?
    \item Do you have a desire to find a partner?
    \item Have you ever felt that your emotional fluctuations are completely out of your control?
    \item Do you often feel inexplicably euphoric or depressed?
    \item Do you believe that financial inadequacies are one of your significant sources of stress?
    \item When you feel mentally exhausted, do you choose to isolate yourself in a room?
    \item Do you have a habit of excessive drinking or smoking?
\end{enumerate}

\vspace{0.5cm}
\noindent \textbf{Note:} By submitting this survey, you agree with our privacy policies.

\end{appendix}

\end{document}